\documentclass[11pt]{article}
\usepackage{amsmath,amssymb,color,graphics,epsfig,cite}
\usepackage[utf8x]{inputenc}

\usepackage{amsfonts}

\newcommand{\be}{\begin{equation}}
\newcommand{\ee}{\end{equation}}
\newcommand{\bea}{\setlength\arraycolsep{2pt} \begin{eqnarray}}
\newcommand{\eea}{\end{eqnarray}}
\newcommand{\nn}{\nonumber}

\newcommand{\w}[1]{\\[0.#1cm]}
\def\ft#1#2{{\textstyle{\frac{\scriptstyle #1}{\scriptstyle #2} } }}
\def\fft#1#2{{\frac{#1}{#2}}}

\def\0{{\sst{(0)}}}
\def\1{{\sst{(1)}}}
\def\2{{\sst{(2)}}}
\def\3{{\sst{(3)}}}
\def\4{{\sst{(4)}}}
\def\5{{\sst{(5)}}}
\def\6{{\sst{(6)}}}
\def\7{{\sst{(7)}}}
\def\8{{\sst{(8)}}}
\def\sst#1{{\scriptscriptstyle #1}}

\def\a{\alpha}
\def\b{\beta}
\def\g{\gamma}

\def\d{\delta}

\def\k{\kappa}
\def\l{\lambda}
\def\L{\Lambda}
\def\m{\mu}
\def\n{\nu}
\def\r{\rho}
\def\s{\sigma}

\def\o{\omega}
\def\O{\Omega}

\begin{document}

\begin{center}
{\Large {\bf $\alpha'$-corrections to Near Extremal Dyonic Strings and Weak Gravity Conjecture  }}

\vspace{20pt}

{\large Liang Ma, Yi Pang,\hbox{*} H. L\"u,}

\vspace{10pt}

{\it Center for Joint Quantum Studies and Department of Physics,\\
School of Science, Tianjin University, Tianjin 300350, China }

\vspace{40pt}

\underline{ABSTRACT}
\end{center}

We construct non-extremal dyonic string solutions in 6D minimal supergravity where the leading
higher derivative corrections arise from either the type IIA string theory compactified on K3
or the heterotic string theory compactified on 4-torus. The thermodynamical quantities and Euclidean actions of the strings are computed. In the near extremal regime, we calculate the force felt by a probe fundamental string in the background of the macroscopic dyonic string with leading $\a'$ corrections. We find that in both the IIA and heterotic setups, away from extremality, the attractive force overwhelms the repulsive force. However, close to extremality, the $\alpha'$ corrections can reduce the attractive force in the isoentropic process, where the charges are fixed. This feature may be used as a new constraint for supergravity models with consistent quantum gravity embedding, in cases where the extremal limit coincides with the BPS limit and the higher derivative corrections do not affect the mass-to-charge ratio. By contrast, the $\alpha'$ corrections can enhance the attractive force in the isothermal or isoenergetic processes.

\vfill{\footnotesize  liangma@tju.edu.cn \ \ \ *pangyi1@tju.edu.cn (corresponding author)
\ \ \ mrhonglu@gmail.com}

\thispagestyle{empty}
\pagebreak



\tableofcontents
\addtocontents{toc}{\protect\setcounter{tocdepth}{2}}

\section{Introduction}

Effective field theories (EFTs) are usually formulated to describe physical processes below a certain ultraviolet cutoff that is much lower than the Planck scale. For this reason, it has long been thought that quantum gravity effects are negligible in the low energy effective theories. Counter intuitively, the weak gravity conjecture (WGC) \cite{Arkani-Hamed:2006emk} and various swampland conjectures \cite{Polchinski:2003bq,Banks:2010zn,Ooguri:2018wrx}\footnote{Some of them were proved in special circumstance, for instance, see \cite{Harlow:2018tng} for the proof of no global symmetry in quantum gravity in asymptotically AdS spacetime.} (See \cite{Palti:2019pca} for a comprehensive review on this subject.) suggest that even at the energy scale well below the Planck scale, there are still nontrivial constraints on the spectrum and symmetry of an effective theory, from the requirement that it should couple to quantum gravity consistently. These constraints are automatically satisfied for models with a string/M theory origin, and can provide useful insights for a generic EFT. The WGC states that there must exist charged particles for which the net force is repulsive. Schematically, this usually implies in the spectrum of the low energy effective theories, there are states with charge-to-mass ratio $q/m>1$. States obeying such a relation can be a fundamental particle much lighter than the Planck scale or macroscopic black holes. Their existence guarantees the extremal Reissner-Nordstrom (RN) black holes admit possible decay channels to release its enormous amount of entropy, thereby avoiding the formation of pathological remnants.

For the time being, studies of the consequences of WGC have been carried out mainly in Einstein-Maxwell or Einstein-Maxwell-dilaton theory, augmented by higher derivative terms. In these frameworks, the charged excitations are the charged black holes, e.g.~the RN black holes, and WGC implies that the higher derivative terms should contribute positively to the charge-to-mass ratio of extremal RN black holes, i.e., for fixed charges, the higher derivative corrections to the mass shift is negative. Thus the structure of the higher derivative terms cannot be arbitrary \cite{Kats:2006xp,Cheung:2018cwt,Cheung:2019cwi,Charles:2019qqt,
Loges:2019jzs,Cremonini:2019wdk,Ma:2020xwi}. Recently, it was also argued that for Einstein-Maxwell theories, the increase of charge-to-mass ratio for extremal RN black holes is related to the positivity of higher derivative corrections to the entropy \cite{Cheung:2018cwt, Cheung:2019cwi}. However, examples from supergravities suggest that the positivity of entropy shift and the increase of charge-to-mass ratio for extremal RN black holes are generically independent of each other \cite{Cano:2019oma,Cano:2019ycn}.\footnote{In fact, positivity of entropy shift caused by higher-dimension operators is a general feature of a unitary EFT, which may not be strong enough to serve as a requirement from quantum gravity.} For instance, when the extremal limit ($T=0$) coincides with the BPS limit, the charge-to-mass ratio is fixed to be 1 in appropriate units, as required by the Poincare supersymmetry algebra and is robust against higher derivative corrections that preserve certain amount of supersymmetry. At the same time, the entropy shift is usually nonvanishing. Clearly, the local supersymmetry alone allows more structures than those descending from string theory. This raises an interesting question: in models where the extremality condition for black holes coincides with the BPS condition, which criteria can substitute for the one on the charge-to-mass ratio as a prerequisite for an EFT to be embeddable in quantum gravity?

In this paper, we will make attempts towards answering this question by studying higher derivative corrections to the near extremal solutions in models with a stringy origin and see if some universal features emerge there. Specifically, the setup we choose is type IIA string compactified on K3. At low energies, the theory is described by $D=6$ ${\cal N}=(1,1)$ supergravity coupled to 20 vector multiplets where the moduli parameterize the $O(4,20)/(O(4)\times O(20))$ coset space. The leading higher derivative corrections arise from the 1-loop string effects and are fourth-order in derivative \cite{Liu:2013dna}. Restricting to the NS-NS sector, the leading higher derivative contribution to the action can be completed using superconformal techniques \cite{Bergshoeff:1985mz,Butter:2016qkx,Butter:2017jqu,Novak:2017wqc, Butter:2018wss}. The theory admits dyonic string solutions that carry both electric and magnetic charges associated with the NS-NS 2-form field. For simplicity, we will focus on general non-extremal static dyonic strings with spherically symmetry in the transverse space. The purely electrically charged string solution has a singular horizon in the extremal limit and will not be considered in this paper. In the extremal limit, the dyonic string solutions becomes BPS and preserve half of the supersymmetry. However, they possess zero entropy. This implies that the usual entropy based argument behind WGC that demands extremal black holes be unstable does not apply here. We thus resort to the original definition of WGC by comparing the attractive force to the repulsive force felt by a test object. After all, the essence of WGC is about comparing the attractive verus repulsive forces \cite{Heidenreich:2019zkl}. In $D=6$ supergravity with stringy origin, the natural test object is the fundamental string that couples to the NS-NS 2-form. The fundamental string itself is also BPS in the sense that its charge is equal to the tension in appropriate units. Using the string action coupled to the background supergravity fields, we can readily see that the probe string feels no net force in the background of the macroscopic extremal black dyonic string, consistent with the fact that there exist multi-center macroscopic dyonic string solutions. Switching to the Einstein frame, this means the total attractive force from exchanging graviton and dilation cancels exactly with the repulsive force mediated by the 2-form. In fact, with the aid of the BPS equations from solving the off-shell Killing spinor equations, one can show that the balance between attractive and repulsive forces persists even when supersymmetric higher derivative terms are turned on. These properties are strong indications that the fundamental string is the right probe in our scenario.

   When heating up the dyonic strings, we can expect that net force should be
attractive as they acquire more neutral energies. This is confirmed by evaluating the static potential imposed on the probe string using the non-extremal dyonic string solutions. Now the interesting question one can ask is what is the sign of higher derivative corrections to the net force, be it positive or negative? Earlier examples of extremal RN black holes indicate that higher derivative interactions compatible with WGC tend to reduce the attractive force. Thus it is conceivable that when the deviation from extremality is small enough, the effects of higher derivative interaction should reduce the attractive force. We will check this explicitly using the near extremal dyonic string solutions in the $D=6$ supergravity model arising from IIA compactified on K3 with 1-loop correction. Since heterotic string on 4-torus is dual to IIA string on K3, one can perform the duality map from IIA to heterotic string arriving at dyonic black string solutions in heterotic string. Similar computations about the net force can thus be done using the fundamental heterotic string as a probe. It should be pointed that these two approaches are not equivalent at low energy since the S-duality involves the Hodge dual of the NS-NS 3-form field strength. In other words, the electric NS-NS string at the strongly-coupled regime becomes magnetic in its dual theory at the weakly-coupled regime.

We find some subtleties arise in the calculations. The $\alpha'$-corrections to the net force depends on how one performs the near extremal expansion in terms of choice of the physical quantities. For the dyonic string solutions with fixed electric and charges, there are 3 choices. We can use temperature $T$, $M-M_0$ or the entropy $S$, where $M_0$ is the ground state energy. These quantities all vanish in the extremal limit; therefore, they can all be used to measure the non-extremality of the black ring. In the near extremal expansion, we fix both the electric and magnetic charges. We find that in both the IIA and heterotic frameworks, up to the leading order in the near extremal expansion, the higher derivative corrections to the net force felt by the probe string can always be repulsive, but only when the expansion is carried out in terms of powers of the entropy, not in terms of temperature or deviation of energy.

This paper is organized as follows. In section 2, we first review some basic properties of the non-extremal dyonic string solutions in 6D minimal supergravity without higher derivative corrections. In section 3 and 4, we embed the 6D minimal supergravity into IIA string on K3 and heterotic string on 4-torus respectively. Utilizing the complete leading $\a'$ corrections to the low energy effective actions based on fields in the NS-NS sector, We obtain the $\a'$ corrections to the non-extremal dyonic string solutions. We then computed the $\a'$ corrected thermodynamic quantities and discuss the force imposed on a probe string by the macroscopic dyonic string near the extremality. We conclude in section 5.  In appendix, we give explicit $\alpha'$ corrections to the non-extremal dyonic black strings.

\section{Dyonic string solutions in $D=6$ 2-derivative supergravity}

The model we consider here is $D=6$ minimal supergravity based on the off-shell Poincare multiplet \cite{Bergshoeff:1985mz,Bergshoeff:2012ax} consisting of  bosonic fields $\{g_{\m\n} , B_{\m\n} , L , V_{\m} , V^{ij}_{\m}, E_{\m}\}$ and fermionic fields $\{\psi^i_{\m}, \varphi^i\}$, with $i = 1, 2$ and $\m,\n$ labeling the 6D coordinate index. In particular, $\{V_{\m} , V^{ij}_{\m}, E_{\m}\}$ are the auxiliary fields needed for the off-shell closure of the supersymmetry algebra. The 2-derivative Lagrangian invariant under the 6D off-shell Poincare supersymmetry for the bosonic sector takes the form
\bea
\label{2driv}
\mathcal{L}_{0}=L (R+L^{-2}\nabla^\mu L\nabla_\mu L-\frac{1}{12}H_{\mu\nu\rho}H^{\mu\nu\rho}+\cdots),\quad H_{\m\n\r}=3\partial_{[\m}B_{\n\r]}\,,
\eea
where the ellipses denote the terms quadratic in the auxiliary fields, which can be set to 0 consistently via field equations. This action is naturally formulated in the string frame and $L$ is usually denoted as $e^{-2\phi}$, where $\phi$ is the dilaton. It describes the low energy dynamics of the NS-NS sector of the string theory compactified on 4-manifolds that preserve 8 supercharges. Later on, we will see that this frame is also convenient for
the construction of supersymmetric higher derivative terms. The theory admits a static black dyonic string solution of the form
\bea
\label{non-BPS Einstein solution}
ds^2_6&=&D(r)(-h(r)dt^2+dx^2)+H_p(r)(\frac{dr^2}{f(r)}+r^2d\Omega^2_3)\,,\cr
B_{(2)}&=&2P\sqrt{1+\frac{\mu}{P}}\omega_{(2)}+\sqrt{1+\frac{\mu}{Q}}A(r)dt\wedge dx\,,
\qquad L=\frac{1}{D(r)H_p(r)},\cr
D(r)&=&A(r)=\frac{r^2}{r^2+Q},\quad
H_p(r)=1+\frac{P}{r^2},\quad  h(r)=f(r)=1-\frac{\mu}{r^2}\,,
\eea
in which the parameters $\m,\, Q,\,P\ge0$ and $\o_{(2)}=-\ft14\cos^2\frac{\theta}{2}d\phi\wedge d\chi$, so $d\o_{(2)}={\rm Vol}(S^3)$. Here the line element on $S^3$, $d\Omega^2_3=\frac{1}{4}(\sigma_3^2+d\Omega^2_2)$, is expressed as a $U(1)$ bundle over a $S^2$ in which $\sigma_3=d\chi-\cos{\theta}d\phi$, $d\Omega^2_2=d\theta^2+\sin^2{\theta}d\phi^2$. The event horizon is located at $r=r_h=\sqrt{\mu}$
The temperature of the black string is defined through the surface gravity $\k$ corresponding to the Killing vector $\xi_t=\partial_t$
\be
\k^2=-\frac{g^{\m\n}\partial_{\m}\xi^2_t\partial_\n\xi^2_t}{4\xi_t^2}\bigg|_{r=r_h},\quad T=\frac{\k}{2\pi}=\frac{1}{2\pi}\frac{\sqrt{\mu }}{\sqrt{\mu +P} \sqrt{\mu +Q}}\,.\label{temperature}
\ee
The entropy density along the $x$-direction is computed using Iyer-Wald formula \cite{Iyer:1994ys}
\be
S=-\frac1{8}\int_{S^3} d\O_3\frac{\partial {\cal L}_0}{\sqrt{-g}\partial R_{\m\n\r\s}}\epsilon_{\m\n}\epsilon_{\r\s}\bigg|_{r=r_h}=\frac{\pi^2}{2} \sqrt{\mu } \sqrt{\mu +P} \sqrt{\mu +Q}\,,
\ee
where the 6D Newton's constant $G_N$ is set to 1 and $\epsilon_{\m\n}$ is the binormal vector of the black string horizon.
The electric and magnetic charges carried by the string are given by
\bea
Q_{\rm e}&=&\frac{1}{16\pi}\int_{S^3}L\star H_{(3)}=\frac{ \pi}{4}  \sqrt{Q} \sqrt{\mu +Q}\,,\\
Q_{\rm m}&=&\frac{1}{16\pi}\int_{S^3}H_{(3)}=\frac{ \pi}{4}  \sqrt{P} \sqrt{\mu +P}\,.
\eea
Accordingly, the electric and magnetic potential can be computed as
\bea
\Phi_{\rm e}&=&B_{tx}|_{r=\infty}-B_{tx}|_{r=r_h}= \sqrt{\frac{Q}{\mu +Q}}\,,\\
\Phi_{\rm m}&=&\widetilde{B}_{tx}|_{r=\infty}-\widetilde{B}_{tx}|_{r=r_h}= \sqrt{\frac{P}{\mu +P}}\,,
\eea
where $\widetilde{B}_{tx}$ is the component of the dual 2-form potential defined via $L\star H_{(3)}=d\widetilde{B}_{(2)}$.

The extremal limit $(T=0)$ is achieved by setting $\mu=0$, which coincides with the BPS
limit.  In this limit, for $PQ\ne 0$, the horizon is located at $r=0$ and the near horizon geometry becomes AdS$_3\times S_3$.  When only $P$ or $Q$ is turned on, the $r=0$ becomes the naked singularity and we shall not consider this situation in this paper. From the first law of thermodynamics, up to a pure constant one can solve for the black string energy density along the $x$-direction
\be
M=\frac{\pi}{4}(P+Q)+\frac{3\pi}{8}\m\,.
\ee
The above expression is also reproduced from the Brown-York surface Hamiltonian \cite{Brown:1992br} and
will be confirmed from the Euclidean action computation. From the thermodynamic quantities listed above,
one can readily see that at fixed electric and magnetic charges, the BPS limit is achieved at $\m=0$ where
\be
M|_{\m=0}=Q_{\rm e}+Q_{\rm m}\,.
\ee
Meanwhile, the temperature $T$ approaches 0. Note that the parameters $Q$ and $P$ stay finite in the $\m\rightarrow 0$ limit for the finiteness of the energy.

The Euclidean action can be calculated using the counter-term extended action
\be
\label{action1}
I_{E}=I_0+I_{\mathrm{GHY}}-I_{c}\,,
\ee
where $I_0$ is the action associated with the 2-derivative Lagrangian \eqref{2driv}, while $I_{\mathrm{GHY}}$ and $I_c$ are
the string frame analogue of Gibbons-Hawking-York boundary term and background subtraction term given below
\be
I_{\mathrm{GHY}}=\frac1{8\pi}\int_{\partial M} d^5x\sqrt{-h}LK,\quad I_c=\frac1{8\pi}\int_{\partial M} d^5x\sqrt{-h}L\bar{K}\,.
\ee
The boundary is located at some large value of the radial coordinate $r=r_c$ which will be taken to infinity at the end of calculation.
$K$ is the trace of the extrinsic curvature of the $r=r_c$ hypersurface embedded in \eqref{non-BPS Einstein solution}. $\bar{K}$ is the trace of the extrinsic curvature of the $r=r_c$ hypersurface embedded in the flat background metric used in the subtraction procedure
\be
d\bar{s}^2_6=D(r_c)(h(r_c)d\tau^2+dx^2)+dR^2+R^2d\Omega^2_3,\quad R^2=r^2H_p(r)\,.
\ee
Substituting the solution \eqref{non-BPS Einstein solution} into the total action \eqref{action1},
we obtain
\be
I_E=\b G\,,\quad \b=T^{-1}\,,\quad G=\frac{\pi}8(\m+2P)=M-TS-\Phi_{\rm e}Q_{\rm e}\,.
\ee
For the convenience of later discussion, we can also express Euclidean action in term of $(T,\,\Phi_{\rm e},\,Q_{\rm m})$. The reason for this choice is that the variation principle requires those quantities being fixed, indicating that the Euclidean is a functional of those variables.
Using the relation
\be
\mu=\frac{64 T^2 Q_{\rm m}^2}{\Phi _{\rm e}^2-1}-\frac{\Phi _{\rm e}^2-1}{4 \pi ^2 T^2},\quad Q=\frac{\Phi _{\rm e}^2}{4 T^2}\left(\frac{1}{\pi ^2}-\frac{256 T^4 Q_{\rm m}^2}{\left(\Phi _{\rm e}^2-1\right){}^2}\right),\quad P=-\frac{64 T^2 Q_{\rm m}^2}{\Phi _{\rm e}^2-1}\,,
\ee
we obtain
\be
I_E=-\frac{\pi}8\beta(\frac{64 T^2 Q_{\rm m}^2}{\Phi _{\rm e}^2-1}+\frac{\Phi _{\rm e}^2-1}{4 \pi ^2 T^2})\,.
\ee
In the next section, we will study the higher derivative corrections to the force acting on a probe fundamental string by the macroscopic
non-extremal dyonic string. As a warm-up, we first examine the force imposed on a probe string by the dyonic string solution \eqref{non-BPS Einstein solution} in the 2-derivative theory. The action of the fundamental string coupled to background supergravity fields takes the form
\be
S_2=-T_2\int d^2\xi \left(\sqrt{-\det\g}+\ft12\epsilon^{ij}\partial_iX^M\partial_jX^NB_{MN}\right)\,,\quad \g_{ij}=g_{MN}\partial_iX^M\partial_jX^N\,,
\ee
where $g_{MN}$ is the target space string frame metric.
To extract the static potential felt by the stirng, we make the static gauge choice
\be
X^{\m}=\xi^{\m}\,,\quad \m=0,\,1\,,
\ee
and consider the transverse coordinate be constant. The potential is then given by
the opposite of the Lagrangian density \cite{Duff:1994an}
\be
{\cal V}=T_2\left(\sqrt{-g_{tt}g_{xx}}-B_{tx}\right)\,.
\ee
Plugging the solution \eqref{non-BPS Einstein solution}, we find
\be
{\cal V}(r)=\frac{T_2r^2}{r^2+Q}\left(\sqrt{1-\frac{\mu}{r}}-\sqrt{1+\frac{{\mu}}{{Q}}}\right)\,.
\ee
Correspondingly, the force is given by
\be
{\cal F}=-\frac{d{\cal V}(r)}{dr}\,.
\ee
It is evident that in the BPS case $\mu=0$, the potential vanishes. In the non-extremal case,
we perform the weak field expansion of the potential in the region where $r\gg {\rm max}\{\sqrt{\mu},\,\sqrt{Q},\,\sqrt{P}\}$. We find
that
\be
\frac{{\cal V}(r)}{T_2}=1-\sqrt{1+\frac{\mu}{Q}}+\frac{\mu}2\frac{1-\sqrt{1+\frac{\mu}{Q}}}{1+\sqrt{1+\frac{\mu}{Q}}}r^{-2}+{\cal O}(r^{-4})\,,
\ee
from which one can see that at leading order the force ${\cal F}=-d{\cal V}(r)/dr$ is attractive.
To separate the gravitational contribution from the scalar contribution,
we replace the string metric by
\be
\label{ES}
g_{MN}=L^{-\ft12}g^E_{MN}\,.
\ee
Then the static potential can be rewritten as
\be
{\cal V}=T_2\left(L^{-\ft12}\sqrt{-g^E_{tt}g^E_{xx}}-B_{tx}\right)\,,
\ee
from which one can clearly distinguish the gravitational contribution from that of the dilaton. In the asymptotic region
where $r\gg {\rm max}\{\sqrt{\mu},\,\sqrt{Q},\,\sqrt{P}\}$, we have
\bea
\sqrt{-g^E_{tt}g^E_{xx}}&=&1-\frac{P+Q+\mu}{2 r^{2}}+{\cal O}(r^{-4})\,,\cr
L^{-\ft12}(r)&=&1+\frac{P-Q}{2r^{2} }+{\cal O}(r^{-4})\,,\cr
B_{tx}&=&\sqrt{1+\frac{\mu}{Q}}-\sqrt{1+\frac{\mu}{Q}}\frac{Q}{r^2}+{\cal O}(r^{-4})\,.
\eea
One should notice that what appears in the leading falloff of the gravitational contribution is different from the ADM mass. The force
mediated by the dilaton can be attractive or repulsive depending on the relative magnitude of $P$ and $Q$.
Also, in the 2-derivative solution, when $P=Q$, the dilaton is constant in the full space-time, which ceases to be the case when
the higher derivative interactions are turned on.

\section{$\alpha'$-corrections from IIA string on K3 }
\label{sec:2aonk3}

When the 6D supergravity model is embedded in IIA string on K3, at low energy the leading higher derivative corrections
arise from 1-loop stringy effects and is fourth-order in derivative \cite{Liu:2013dna}. The fourth-order interactions can be completed using superconformal techniques \cite{Butter:2017jqu,Novak:2017wqc}. They correspond to a particular combination of the supersymmetric Gauss-Bonnet term and Riemann tensor squared of the following form
\bea
\label{4derivaction}
\mathcal{L}_{\mathrm{GB}}&=&R_{\mu\nu\rho\sigma}R^{\mu\nu\rho\sigma}-4R_{\mu\nu}R^{\mu\nu}
+R^2+\frac{1}{6}RH^2
-R^{\mu\nu}H^2_{\mu\nu}+\frac{1}{2}R_{\mu\nu\rho\sigma}H^{\mu\nu\lambda}H^{\rho\sigma}_{\ \ \lambda}\cr
&&+\frac{5}{24}H^4+\frac{1}{144}\left(H^2\right)^2-\frac{1}{8}\left(H^2_{\mu\nu}\right)^2
+\frac{1}{4}\epsilon^{\mu\nu\rho\sigma\lambda\tau}B_{\mu\nu}R_{\rho\sigma\ \beta}^{\ \ \alpha}(\omega_+)R_{\lambda\tau\ \alpha}^{\ \ \beta}(\omega_+)\,,\nn\\
\mathcal{L}_{\mathrm{Riem}^2}&=&R_{\mu\nu\alpha\beta}(\omega_-)R^{\mu\nu\alpha\beta}(\omega_-)
+\frac{1}{4}\epsilon^{\mu\nu\rho\sigma\lambda\tau}B_{\mu\nu}R_{\rho\sigma\ \beta}^{\ \ \alpha}(\omega_-)R_{\lambda\tau\ \alpha}^{\ \ \beta}(\omega_-)\,,
\eea
where in our convention $\sqrt{-g}\epsilon^{012345}=-1$ and terms quadratic in auxiliary fields are omitted.  Here $R_{\mu\nu\ \beta}^{\ \ \alpha}(\omega_{\pm})$ is the curvature defined with respect to the torsionful spin connection $\omega^{\alpha}_{\pm\mu\beta}$
\bea
R_{\mu\nu\ \beta}^{\ \ \alpha}(\omega_{\pm})=\partial_{\mu}\omega^{\alpha}_{\pm\nu\beta}+
\omega^{\alpha}_{\pm\mu\gamma}\omega^{\gamma}_{\pm\nu\beta}-(\mu\leftrightarrow\nu)\,,
\ \ \ \omega^{\alpha}_{\pm\mu\beta}=\omega^{\alpha}_{\mu\beta}\pm\ft{1}{2}H_{\mu\ \beta}^{\ \alpha}\,,
\eea
and the shorthand notations for various contractions of $H_{\mu\nu\rho}$ are defined as
\bea
H^2=H_{\mu\nu\rho}H^{\mu\nu\rho},\ \ \ H^2_{\mu\nu}=H_{\mu\rho\sigma}
H_{\nu}^{\ \rho\sigma},\ \ \ H^4=H_{\mu\nu\sigma}H_{\rho\lambda}^{\ \ \sigma}H^{\mu\rho\delta}H^{\nu\lambda}_{\ \ \ \delta}\,.
\eea
Each of these two superinvariants preserves 8 supercharges. However, a special combination of them with equal coefficients preserves 16 supercharges, as this is the right combination that appears in the IIA string amplitude calculation \cite{Liu:2013dna} and satisfies the $S$-duality relation with the tree-level $\a'$ correction of heterotic string compactified on $T^4$. Specifically, including the leading $\a'$
correction, the model now takes the form
\be
\label{tot}
\mathcal{L}_{LR+R^2}=\mathcal{L}_{0}+\frac{\a'}{16}(\mathcal{L}_{\mathrm{GB}}+\mathcal{L}_{\mathrm{Riem}^2})\,.
\ee
We now construct the dyonic string solutions in the model with $\a'$ corrections. It turns out that
the ansatz that accommodates the dyonic string needs a slight extension. An off-diagonal term proportional to
$dtdx$ terms must be included. Thus the full ansatz takes the form
\bea
\label{non-BPS-alpha}
ds^2_6=&&D(r)\left(-h(r)dt^2+dx^2+2\omega dtdx\right)+H_p(r)\left(\frac{dr^2}{f(r)}+r^2d\Omega^2_3\right),\cr
B_{(2)}=&&2P\sqrt{1+\frac{\mu}{P}}\omega_{(2)}+\sqrt{1+\frac{\mu}{Q}}A(r)dt\wedge dx\,,\quad L=L(r)\,,
\eea
where the functions $L(r),\,D(r),\,A(r),\,f(r),\,h(r),\,\omega(r)$ can be written as the leading order
solution \eqref{non-BPS Einstein solution} plus perturbations
\be
\label{non BPS perturbation}
L=L_0+\delta L,\ \ \ D=D_0+\delta D,\ \ \ A=D_0+\delta A,\ \ \ f=f_0+\delta f,\ \ \ h=f_0+\delta h,\ \ \ \omega=\delta\omega\,,
\ee
where we use subscript ``0" to label the solution in the 2-derivative theory and the perturbations are proportional $\a'$.
The field equations involving $\a'$ corrections were given in \cite{Butter:2018wss}. We substitute the ansatz \eqref{non BPS perturbation}
to the equations of motion and keep only term linear in $\a'$. The resulting equations are a set of coupled linear ODE for the perturbations. The generic solution is given in appendix (A.1). It is characterised by 8 integration constants $C_i,\,i=1,\cdots,8$. Requiring the solution be regular on the horizon and asymptotic to Minkowski fixes the parameters to be
\bea
\label{solution 1}
&&\{C_1,\,C_2,\,C_3,\,C_4,\,C_5,\,C_6,\,C_7,\,C_8\}\cr
&=&\{0,\frac{\mu ^2 (3 \mu +2 Q)\a' }{4 (\mu +Q)^2},\frac{3 \mu ^2 (3 \mu +2 Q) \a'}{4 Q (\mu +Q)^2 (\mu +2 Q)}, 0,0,0,0,0\}\,.
\eea
Different from the solution in 2-derivative theory, the $dtdx$ component no longer vanishes but acquires a $\a'$ correction
\be
\delta\omega=\frac{\mu  \sqrt{P} \sqrt{\mu +P} }{2 r^2 (P+r^2) (Q+r^2)}\a'\,.
\ee
It is generated by the parity violating $B\wedge R\wedge R$ terms in the action. The finite temperature also plays a role in its presence, as $\d \o=0$ at $\m=0$. The horizon is located at the zero of
\be
(g_{tt}g_{xx}-g_{tx}^2)|_{r=r_h}=0\,.
\ee
Because $g_{tx}$ is of the order of $\a'^2$,  to the first order in $\a'$ one can simply compute $r_h$ from the zero of $g_{tt}$ as $g_{xx}$ always stays positive. The result is given by
\be
r_h\rightarrow \sqrt{\mu}+\alpha'\delta r\,,\quad   \delta r=\frac{\sqrt{\mu } \left(\mu  (\mu +P)-4 \mu  Q-4 Q^2\right)}{16 Q (\mu +Q)^2}
-\frac{\mu ^{5/2} (\mu +P)}{16 Q^2 (\mu +Q)^2}\log \left(1+\frac{Q}{\mu }\right).
\ee
At $r=r_h$, the line element in the $dt$ and $dx$ direction becomes degenerate as
\be
ds^2|_{r=r_h}=\frac1{g_{xx}}(g_{tx}dt+g_{xx}dx)^2|_{r=r_h}\,.
\ee
Thus there is a velocity of order $\a'$ in the $x$-direction given by
\be
v_x=-\frac{g_{tx}}{g_{xx}}|_{r=r_h}\,.
\ee
However, since $g_{tx}$ falls off as $r^{-6}$ near the infinity, the solution carries no linear momentum measured at infinity. This is verified by evaluating the conserved charge associated with Killing vector $\partial_x$ and by the first law of thermodynamics studied below in which linear momentum does not appear. It is worth noting that now the Killing vector that is null on the shifted horizon $r_h$ is $\xi=\partial_t + v_x \partial_x$. We can obtain the surface gravity and hence the temperature using the formula (\ref{temperature}).

Using the solution, one can compute the $\a'$ corrected thermodynamical quantities
\bea
\label{corrected1}
M&=&\frac{ 3\pi  }{8}\mu+\frac{ \pi  }{4}(Q+P)-\frac{3 \pi  \mu ^2 (3 \mu +2 Q) }{32 (\mu +Q)^2 (\mu +2 Q)}\a',\cr
T&=&\frac{1}{2\pi}\frac{\sqrt{\mu }}{\sqrt{\mu +P} \sqrt{\mu +Q}}-\frac{\sqrt{\mu } Q (5 \mu +4 Q) }{4 \pi  \sqrt{\mu +P} (\mu +Q)^{5/2} (\mu +2 Q)}\a',\cr
S&=&\frac{1}{2} \pi ^2 \sqrt{\mu } \sqrt{\mu +P} \sqrt{\mu +Q}+\frac{\pi ^2 \sqrt{\mu }  \sqrt{\mu +P} Q(5 \mu +4 Q) }{4 (\mu +Q)^{3/2} (\mu +2 Q)}\a',\cr
Q_{\rm e}&=&\frac{ \pi}{4}  \sqrt{Q} \sqrt{\mu +Q},\quad\Phi_{\rm e}=\sqrt{\frac{Q}{\mu +Q}}+\frac{\mu  \sqrt{Q} (5 \mu +4 Q) }{2 (\mu +Q)^{5/2} (\mu +2 Q)}\a',\cr
Q_{\rm m}&=&\frac{ \pi}{4}  \sqrt{P} \sqrt{\mu +P},\quad\Phi_{\rm m}=\sqrt{\frac{P}{\mu +P}}\,.
\eea
These satisfy the first law of thermodynamics up to ${\cal O}(\a'^2)$
\be
dM-TdS-\Phi_{\rm e}dQ_{\rm e}-\Phi_{\rm m}dQ_{\rm m}=\mathcal{O}( \a'^2).
\ee
Different from the 2-derivative theory, the entropy is now computed using Iyer-Wald formula applied to the total
Lagrangian \eqref{tot}. The electric charge is computed from
\be
Q_{\rm e}=\frac{1}{16\pi}\int_{S^3}\star(L H_{(3)}+\frac{\a'}{16}K_{(3)})\,,
\ee
where the components of the 3-form $K_{(3)}$ is given by
\bea
&&K^{\m\n\l}=12R^{\r[\m}H_{\r}{}^{\n\l]}-2RH^{\m\n\l}-\ft16H^2H^{\m\n\l}
-2H^{[\m}{}_{\r\s}H^{2\n|\r|,\l]\s}+4\square H^{\m\n\l}\nn\\
&&\qquad+6(R_{\r\s}{}^{[\m\n}(\o_+)\star H^{\l]\r\s}-R_{\r\s}{}^{[\m\n}(\o_-)\star H^{\l]\r\s})+2\star{\rm CS}^{\m\n\l}(\o_+)+2\star{\rm CS}^{\m\n\l}(\o_-)\,,
\eea
where the components of the Chern-Simons term is defined via
\be
\partial_{[\m}{\rm CS}_{\n\r\l]}(\o_\pm)=\ft32R_{[\m\n\ b}^{\ \ \ a}(\o_\pm)R_{\r\l\ a]}^{\ \ \ b}(\o_\pm)\,.
\ee
Since the $B$-field equation is
\be
d\star(L H_{(3)}+\ft{\a'}{16}K_{(3)})=0\,,
\ee
we can again define the magnetic dual 2-form $\widetilde{B}_{(2)}$ as $d\widetilde{B}_{(2)}=\star(L H_{(3)}+\frac{\a'}{16}K_{(3)})$, from which the
magnetic potential can be calculated. Using the same method as before, we obtain the Euclidean action
\bea
I_{E}&=&\beta\left[\frac{\pi}8(\m+2P)-\frac{\pi  \mu  (9 \mu +8 Q) }{32 (\mu +Q)^2}\a'\right]=\beta(M-TS-\Phi _{\rm e}Q_{\rm e})\,,\nn\\
&=&-\beta\left[\frac{\pi }{8}\left(\frac{64 T^2 Q_{\rm m}^2}{\Phi _{\rm e}^2-1}+\frac{\Phi _{\rm e}^2-1}{4 \pi ^2 T^2}\right)+\frac{\pi}{32} \left(\Phi _{\rm e}^2-9\right) \left(\Phi _{\rm e}^2-1\right)\a'\right].
\eea
Using the $\a'$-corrected thermodynamical quantities \eqref{corrected1}, one can check that
\be
-\frac{\pi }{8}\left(\frac{64 T^2 Q_{\rm m}^2}{\Phi _{\rm e}^2-1}+\frac{\Phi _{\rm e}^2-1}{4 \pi ^2 T^2}\right)=\frac{\pi }{8}(\m+2P)+{\cal O}(\a'^2)\,,
\ee
thus confirming the result of \cite{Reall:2019sah}. In other words, for the action \eqref{tot},
up to first order in $\a'$, in terms of variables $T,\,\Phi _{\rm e},\,Q_{\rm m}$
the $\a'-$corrected on-shell action can be expressed the action
\be
\label{HR}
I_E=I_{E0}(T,\Phi _{\rm e},Q_{\rm m})+I_{\rm hd}(T,\Phi _{\rm e},Q_{\rm m})\,,
\ee
where functional form of $I_{E0}$ is the same as the one derived from evaluating the 2-derivative solution
on the 2-derivative action. $I_{\rm hd}$ is linear in $\a'$ and can be obtained by simply evaluating the uncorrected solution in the higher
derivative part of the action. From \eqref{HR}, one can also derive that given a set of conserved charges $(M,\,Q_{\rm e},\,Q_{\rm m})$
the leading correction to the entropy can be readily computed from
\be
\label{ds1}
\delta S(M,\,Q_{\rm e},\,Q_{\rm m})=-I_{\rm hd}\,.
\ee
The above formula can be derived from combination of two equations below
\bea
\frac{\partial I_{\rm hd}}{\partial \a'}\big|_{(T,\,\Phi _{\rm e},\,Q_{\rm m})}&=&\frac{\partial I_E}{\partial \a'}\big|_{(T,\,\Phi _{\rm e},\,Q_{\rm m})}\cr
&=&-\frac{\partial S}{\partial\a'}\big|_{(T,\,\Phi _{\rm e},\,Q_{\rm m})}+\beta\frac{\partial M}{\partial \a'}\big|_{(T,\,\Phi _{\rm e},\,Q_{\rm m})}-\beta\Phi _{\rm e}\frac{\partial Q_{\rm e}}{\partial\a' }\big|_{(T,\,\Phi _{\rm e},\,Q_{\rm m})}\,,\nn\\
\frac{\partial S}{\partial\a'}\big|_{(T,\,\Phi _{\rm e},\,Q_{\rm m})}&=&\frac{\partial S}{\partial M}\big|_{(Q_{\rm e},\,Q_{\rm m},\,\a')}
\frac{\partial M}{\partial \a'}\big|_{(T,\,\Phi_{\rm e},\,Q_{\rm m})}+\frac{\partial S}{\partial Q_{\rm e}}\big|_{(M,\,Q_{\rm m},\,\a')}
\frac{\partial Q_{\rm e}}{\partial \a'}\big|_{(T,\,\Phi_{\rm e},\,Q_{\rm m})}\nn\\
&+&\frac{\partial S}{\partial\a'}\big|_{(M,\,Q _{\rm e},\,Q_{\rm m})}\,,
\eea
together with the first law of thermodynamics. From (39,40) and \eqref{ds1}, one can see the $\a'$-correction to the entropy is positive for positive $\alpha'$:
\be
\delta S(M,\,Q_{\rm e},\,Q_{\rm m})=\frac{\pi \beta \mu  (9 \mu +8 Q) }{32 (\mu +Q)^2}\a'=\frac{\pi ^2 \sqrt{\mu } \sqrt{\mu +P} (9 \mu +8 Q)}{16 (\mu +Q)^{3/2}}\a'\,.
\ee
It is clear that in the extemal limit $(\mu=0)$, we have $\delta S = {\cal O}(\alpha'^2)
=\delta M$ for fixed electric and magnetic charges, even though $\alpha'$ correction enters the extremal dyonic string. The balance is achieved by precise cancelation of the contributions from the pure matter sector and gravity sector (including non-minimally coupled matter). Indeed we verified that the gravity sector of the $\alpha'$ correction contributes positively to $\delta M$, whilst the pure matter sector contributes negatively, confirming the conjectured made in \cite{Ma:2020xwi}.

We therefore need to examine $\delta M$ in the near extemal limit. Unlike the extremal case where the solutions are specified by the charges only, we now need to fix an extra parameter for the nonextremality.  The only natural choice is either $T$ or $S$, both of which vanish in the extremal limit. Since $(\mu,Q,P)$ are not directly physical quantities, we can require that they also acquire $\alpha'$ correction, with
\be
\mu \rightarrow \mu + \alpha' \delta \mu,\qquad
Q\rightarrow Q + \alpha' \delta Q\,,\qquad
P\rightarrow P + \alpha' \delta P\,,
\ee
such that the charges $(Q_{\rm e},Q_{\rm m})$ and temperature $T$ are the same under the $\alpha'$ correction. We find
\bea
\delta Q &=& -\frac{\mu  Q^2 (\mu +2 P) (5 \mu +4 Q)}{(\mu +Q)^2 (\mu +2 Q) \left(4 P Q-\mu ^2\right)}\,,\qquad \delta P= -\frac{\mu  P Q (5 \mu +4 Q)}{(\mu +Q)^2 \left(4 P Q-\mu ^2\right)}\,,\cr
\delta \mu &=& \frac{\mu  Q (\mu +2 P) (5 \mu +4 Q)}{(\mu +Q)^2 \left(4 P Q-\mu ^2\right)}\,.
\eea
Consequently, we find that
\be
\delta M= \alpha' \Big(4\pi^2 Q_{\rm m} T^2 + {\cal O}(T^3)\Big)>0\,.
\ee
An alternative approach is to fix the charges and the entropy which becomes nonvanishing but small in the near-extremal limit.  In this case, we have negative contribution from the $\alpha'$ correction
\be
\delta M= \alpha' \Big(-\fft{S^2}{64Q_{\rm e}^2 Q_{\rm m}} + {\cal O}(S^3)\Big)<0\,.
\ee
The fact that we get opposite sign using $T$ and $S$ as the perturbative variables may originate from the fact that the $\alpha'$ contributions to $T$ and $S$ in \eqref{corrected1} are opposite in sign.  Thus we see that $\delta M$ is not always negative under the $\alpha'$ correction, depending on whether it is an isothermal or isoentropic process. Of course, as was explained earlier, $\delta M$ is not the proper indicator for the change of the total attractive force when a scalar field is also involved.

We now investigate the force imposed on a probe string by dyonic string solution with higher derivative corrections. We adopt the same action for the probe string. The fact that there is no correction to the fundamental string action is an advantage of working in the framework of string theory. The loop corrections (controlled by $\a'$) in the string world-sheet sigma model contribute to the beta function of background supergravity fields. Setting beta functions to 0 yields the $\a'$ corrections to the supergravity field equations. Thus once the supergravity fields are on-shell which is the case here, beta functions on the string world-sheet sigma model vanish and there is no loop ($\a'$) correction to the string action itself. As we see presently, the force calculated using the probe action in the $\alpha'$-corrected background vanishes in the extremal limit (BPS and static), thereby confirming this string paradigm. However, in a general framework without string theory origin, one would have to consider higher order corrections in the probe string action.

Therefore, the static potential can be rewritten as
\be
{\cal V}=T_2\left(\sqrt{-g_{tt}g_{xx}+g_{tx}^2}-B_{tx}\right)\,.
\ee
In terms of the fields in Einstein frame
\be
{\cal V}=T_2\left(L^{-\ft12}\sqrt{-g^E_{tt}g^E_{xx}+g^{E2}_{tx}}-B_{tx}\right)\,.
\ee
The solution to the off-shell Killing spinor equation requires \cite{Pang:2019qwq}
\be
B_{tx}=\sqrt{-g_{tt}g_{xx}}\,.
\ee
Thus for static BPS solution, the attractive force always balances the repulsive force.

Substituting the $\a'$-corrected dyonic string solution and expanding the potential in the region $r\gg {\rm max}\{\sqrt{\mu},\,\sqrt{Q},\,\sqrt{P}\}$
we find that the static potential takes the form
\bea
\frac{{\cal V}(r)}{T_2}&=&1-\sqrt{1+\frac{\mu}{Q}}-\fft{\Sigma}{r^2}+{\cal O}(r^{-3})\,,\nn\\
\Sigma &=&\frac{\mu }{2}+Q \left(1-\sqrt{1+\frac{\mu}{Q}}\right)-\frac{3 \left(3 \mu+2 Q\right)\mu ^2}{8 (\mu +Q)^2 (\mu +2 Q)}\alpha'\,.
\eea
Since $g_{tx}\sim r^{-6}$ in the large $r$ expansion, it does not contribute to $\Sigma$. In the Einstein, we
can separate the gravitational force from that of scalar exchange. The large $r$ expansions of the fields in Einstein frame
are given by
\bea
\sqrt{-g^E_{tt}g^E_{xx}}&=&1-\frac{\s_g}{ r^2}+{\cal O}(r^{-4})\,,\quad \s_g=\frac12\left(P+Q+\mu-\frac{3(3 \mu+2 Q)\mu ^2 }{8(Q+\m)^2(2Q+\m)}\a'\right)\,,\cr
L^{-\ft12}(r)&=&1-\frac{\s_L}{ r^2}+{\cal O}(r^{-4})\,,\quad\s_L=\frac12\left(Q-P-\frac{3(3 \mu +2 Q)\mu ^2}{8(Q+\m)^2(2Q+\m)}\a'\right)\,,
\cr
B_{tx}&=&\sqrt{1+\frac{\mu}{Q}}-\frac{\s_B}{r^{2}}+{\cal O}(r^{-4})\,,\quad \s_B=Q\sqrt{1+\frac{\mu}{Q}}\,.
\eea
From (53) and (54), one can see that the total force is proportional to the sum of three contributions, gravity, scalar and electric charges:
\be
\Sigma=\s_g+\s_L-\s_B\,.
\ee
It is clear that if the $\alpha'$ correction $\delta \Sigma$ is positive, it enhances the attractive force, while a negative $\delta \Sigma$ is repulsive.

However, in the presence of $\a'$-correction, one needs to act carefully when discussing genuine $\a'$ correction to the static potential expressed in terms of $(\m,\,P,\,Q)$ which are not physical quantities and can be redefined with $\a'$ dependent terms. However, the relation among physical quantities should not depend on the reparameterization. We thus need to recast $\Sigma$ in terms of physical quantities in order to discuss the $\a'$-correction to the static force. There seem to be 3 equally reasonable choices. We would like to express $\Sigma$ in terms of $(T,\,Q_{\rm e},\,Q_{\rm m})$, $(M,\,Q_{\rm e},\,Q_{\rm m})$ or $(S,\,Q_{\rm e},\,Q_{\rm m})$. They correspond to the inclusion of the higher derivative interactions in
3 different physical processes, all with fixed electric and magnetic charges: 1) isothermal process; 2) isoenergetic process; 3) isoentropic process. We can first solve for parameters $Q$ and $P$ from \eqref{corrected1}
\be
\label{QP}
Q=\frac{\sqrt{\pi ^2 \mu ^2+64 Q_{\rm e}^2}-\pi  \mu }{2 \pi }\,,\quad P=\frac{\sqrt{\pi ^2 \mu ^2+64 Q_{\rm m}^2}-\pi  \mu }{2 \pi }\,.
\ee
Then we need to solve for $\m$ in terms of the physical quantities, which in general, cannot be done explicitly due to the complicated structure of the $\a'$-corrections to the thermodynamic quantities \eqref{corrected1}. However, we can solve $\m$ in the near extremal
limit systematically. For the purpose of understanding weak gravity conjecture in supergravity models, it is also meaningful to consider
the near extremal expansion of $\Sigma$.
\begin{description}
    \item[1)] In the first case, we express $\mu$ using $(T\,,Q_{\rm e}\,,Q_{\rm m})$ about $T=0$. We find that
    \be
    \mu=c_0T^2+c_2T^4+{\cal O}(T^6)\,,
    \ee
    where
    \bea
    c_0&=&64 Q_{\rm e} Q_{\rm m}+32 \pi  Q_{\rm m} \alpha'\,,\cr
    c_2&=&512\pi Q_{\rm e}Q_{\rm m}(Q_{\rm e}+Q_{\rm m})+128\pi^2(4Q_{\rm e}+Q_{\rm m})Q_{\rm m}\a'\,.
    \eea
    We can also perform a low temperature expansion of $\Sigma$  at fixed electric and magnetic charges
     \be
     \Sigma=128 \pi  Q_{\rm e} Q_{\rm m}^2 T^4+32 \pi ^2 Q_{\rm m}^2 T^4 \alpha'+{\cal O}(T^6)\,,
     \ee
     from which we see that the higher derivative corrections seem to enforce the attractive force.
     In terms of the leading falloff coefficients of various fields in the Einstein frame, up to first order in $\alpha$, we have
   \bea
   \s_g&=&\frac{2 (Q_{\rm e}+Q_{\rm m})}{\pi }+64 \pi Q_{\rm e} Q_{\rm m}  \left(  Q_{\rm e}+  Q_{\rm m}\right)T^4
   +16\pi^2  Q_{\rm m}\left(4 Q_{\rm e} + Q_{\rm m}\right)T^4\a'+{\cal O}(T^6)\,,\cr
   \s_L&=&\frac{2 (Q_{\rm e}-Q_{\rm m})}{\pi }-64\pi Q_{\rm e} Q_{\rm m} \left( Q_{\rm e}- Q_{\rm m}\right) T^4
   +16\pi^2  Q_{\rm m} \left(  Q_{\rm m}-4 Q_{\rm e} \right)T^4\a'+{\cal O}(T^6)\,,\cr
   \s_B&=&\frac{4 Q_{\rm e}}{\pi}+{\cal O}(T^6)\,,
   \eea
   from which we can see that at fixed electric and magnetic charges, the $\a'$ correction tends to increase the gravitational force
   when the system deviates from extremality by a small temperature. The sign of the $\a'$ correction to the scalar force depends on the relative magnitude between $Q_{\rm e}$ and $Q_{\rm m}$. However, their sum is always attractive.

    \item[2)]  In the second case, we express $\mu$ using $(M\,,Q_{\rm e}\,,Q_{\rm m})$ near the BPS limit in which $M$ approaches $M_0=Q_{\rm e}+Q_{\rm m}$.
    Thus we define the expansion parameter as $\delta M=M-M_0$. $\m$ can be expanded in terms of powers of $\delta M$
    \be
    \mu=c_1\delta m+c_2(\delta m)^2+{\cal O}(\delta m)^3\,,
    \ee
    where
    \be
    c_1=\frac{8}\pi\,,\quad c_2=\frac{-4 Q_{\rm e}^2-4 Q_{\rm e}Q_{\rm m}+3 \pi Q_{\rm m}\alpha' }{\pi Q_{\rm e}^2Q_{\rm m}}\,.
    \ee
    Meanwhile, the leading coefficient in the potential acquires an expansion
    \bea
  \Sigma&=& \frac{\pi c_1^2 \left(4    Q_{\rm e}-3 \pi   \alpha' \right)}{128 Q_{\rm e}^2}\delta m^2+{\cal O}(\delta m^3)\,,\cr
  &=& \frac{2 \delta m^2}{\pi  Q_{\rm e}}-\frac{3 \delta m^2 \a' }{2 Q_{\rm e}^2}+{\cal O}(\delta m^3)\,.
    \eea
    from which one can see that the higher derivative corrections appear to reduce the attractive force.
     In terms of the leading falloff coefficients of various fields in Einstein frame, up to first order in $\a'$
     we have
   \bea
   \s_g&=&\frac{2 (Q_{\rm e}+Q_{\rm m})}{\pi }+\frac{Q_{\rm e}+Q_{\rm m}}{Q_{\rm e}Q_{\rm m}}\frac{\delta m^2}{\pi}
   -\frac{3 \delta m^2 }{4 Q_{\rm e}^2}\a'+{\cal O}(\delta m^3)\,,\cr
   \s_L&=&\frac{2 (Q_{\rm e}-Q_{\rm m})}{\pi }+\frac{Q_{\rm m}-Q_{\rm e}}{Q_{\rm e}Q_{\rm m}}\frac{\delta m^2}{\pi}
   -\frac{3 \delta m^2 }{4 Q_{\rm e}^2}\a'+{\cal O}(\delta m^3)\,,\cr
   \s_B&=&\frac{4 Q_{\rm e}}{\pi}+{\cal O}(\delta m^3)\,.
   \eea
   In this case we see that at fixed electric and magnetic charges, the $\a'$ correction tends to decrease both the gravitational force
  and the scalar force when the energy of the system increases slightly from the extremal value.
  \item[3)]  In the last case, we express $\mu$ using $(S\,,Q_{\rm e}\,,Q_{\rm m})$ near the BPS limit in which $S\rightarrow 0$.
    To distinguish with the entropy in the thermodynamics, we use $s$ to denote the small entropy expansion parameter. We find that
     $\mu$ can be expanded as
    \be
    \mu=c_0s^2+c_2s^4+{\cal O}(s^6)\,,
    \ee
    where up to first order in $\alpha$
    \be
    c_0=\frac{1}{4 \pi ^2 Q_{\rm e} Q_{\rm m}}-\frac{\alpha' }{8 \pi  Q_{\rm e}^2 Q_{\rm m}}\,,\quad
     c_2= -\frac{Q_{\rm e}+Q_{\rm m}}{128 \pi ^3 Q_{\rm e}^3 Q_{\rm m}^3}+\frac{(4 Q_{\rm e}+7 Q_{\rm m})}{512 \pi ^2 Q_{\rm e}^4 Q_{\rm m}^3}\a'
    \ee
    Meanwhile, the leading coefficient in expansion of the potential acquires a small entropy expansion
    \bea
   \Sigma&=&\frac{\pi c_0^2\left(4   Q_{\rm e}-3 \pi    \alpha' \right)}{128 Q_e^2}s^4+{\cal O}(s^6)\nn\\
   &=&\frac{s^4}{512 \pi ^3 Q_{\rm e}^3 Q_{\rm m}^2}-\frac{7 s^4 }{2048 \pi ^2 Q_{\rm e}^4 Q_{\rm m}^2}\alpha'+{\cal O}(s^6)
    \eea
    from which one can see that the higher derivative corrections appear to reduce the attractive force.
    In terms of the leading falloff coefficients of various fields in Einstein frame, up to first order in $\a'$
    we have
   \bea
   \s_g&=&\frac{2 (Q_{\rm e}+Q_{\rm m})}{\pi }+\frac{Q_{\rm e}+Q_{\rm m}}{1024 \pi ^3 Q_{\rm e}^3 Q_{\rm m}^3}s^4
   -\frac{4 Q_{\rm e}+7 Q_{\rm m}}{4096 \pi ^2 Q_{\rm e}^4 Q_{\rm m}^3}s^4\a'+{\cal O}(s^6)\,,\cr
   \s_L&=&\frac{2 (Q_{\rm e}-Q_{\rm m})}{\pi }+\frac{Q_{\rm m}-Q_{\rm e}}{1024 \pi ^3 Q_{\rm e}^3 Q_{\rm m}^3}s^4
   -\frac{7 Q_{\rm m}-4 Q_{\rm e}}{4096 \pi ^2 Q_{\rm e}^4 Q_{\rm m}^3}s^4\a'+{\cal O}(s^6)\,,\cr
   \s_B&=&\frac{4 Q_{\rm e}}{\pi}+{\cal O}(s^6)\,.
   \eea
   In this case we see that at fixed electric and magnetic charges, the $\a'$ correction reduces the gravitational force
  when the entropy of the system increases slightly from 0. The sign of the $\a'$ correction to the scalar force depends on the relative magnitude between $Q_{\rm e}$ and $Q_{\rm m}$. However, their sum is always repulsive.
    \end{description}

\section{$\alpha'$-corrections from heterotic string on 4-torus}

The 6D (1,1) theory can also come from compactification of heterotic string on 4-torus. $\a'$ corrections to the heterotic string effective action has been studied long time ago \cite{Gross:1986mw,Bergshoeff:1989de}.
The leading $\a'$ correction appears in the tree level low energy effective action and is also
fourth-order in derivative. Reducing the theory on $T^4$ and retaining only fields in the 6D minimal Poincare
supergravity multiplet, one arrives at the action below
\be
\label{het}
{\cal L}_{LR+LR^2}=\sqrt{-g}L\left(R+L^{-2}\nabla^\mu L\nabla_\mu L-\frac{1}{12}\widetilde{H}_{\mu\nu\rho}\widetilde{H}^{\mu\nu\rho}+\frac{\alpha'}{8}R_{\mu\nu\alpha\beta}(\omega_+)R^{\mu\nu\alpha\beta}(\omega_+)\right)
\ee
where the 3-form field strength $\tilde{H}_{\m\n\r}$ includes the Lorentz Chern-Simons term
\bea
\widetilde{H}_{(3)}=dB_{(2)}+\frac{1}{4}\alpha'{\rm CS}_{(3)}(\omega_+)\,,
\eea
such that it obeys a deformed Bianchi identity
\bea
\label{non-trivial Bianchi}
d\widetilde{H}_{(3)}=\frac{1}{4}\alpha'R^a_{\ b}(\omega_+)\wedge R^b_{\ a}(\omega_+).
\eea
String dualities imply that the theory above is connected to the one from IIA via $S$-duality.
Up to first order in $\a'$, it has been suggested that the following duality equations still hold \cite{Liu:2013dna}
\be
L^{\rm \sst{IIA}}\star H_{(3)}^{\rm IIA}=\widetilde{H}_{(3)}^{\rm het},\quad L^{\rm IIA}g_{\mu\nu}^{\rm IIA}=g_{\mu\nu}^{\rm het},\quad L^{\rm IIA}=\frac{1}{{L^{\rm het}}}\,.
\ee
Since the string frame metric is related to the Einstein frame metric via \eqref{ES}, the $S$-duality implies
the Einstein frame metrics on both sides should be identical. Tentatively, one can take the solution in the
IIA side and obtain a solution in the heterotic side using the relation above. We checked that the resulting
field configurations indeed solves the equations of motion derived from \eqref{het}.
After redefining the parameters and performing a change of radial coordinate
\bea
&&Q\rightarrow P,\quad P\rightarrow Q,\quad  t\rightarrow-t\,,\quad r\rightarrow r+\a' \delta r\,,\cr
&&\delta r=\frac{r^3 }{16 Q (\mu +Q)^2 (\mu +2 Q)(Q+r^2)^2}\Big(\mu ^2 (3 \mu ^2+2 P (\mu +2 Q)+14 Q^2+19 \mu  Q)\cr
&&-\frac{Q (\mu ^4 P-2 \mu ^2 Q^2 (P-9 \mu )+\mu ^3 Q (4 \mu +P)+56 \mu ^2 Q^3+56 \mu  Q^4+16 Q^5)}{r^4}\cr
&&-\frac{\mu ^2\left(P (\mu ^2-6 Q^2-\mu  Q)-Q (7 \mu ^2+30 Q^2+41 \mu  Q )\right)}{r^2}
\Big)\,,
\eea
we can bring the dyonic string solution in the theory \eqref{het} into the form \eqref{non-BPS-alpha}.
The explicit form of the perturbations are given in appendix (A.2). When solving for $B_{(2)}$ from
$\widetilde{H}_{(3)}$, we had encounter a term from ${\rm CS}_{(3)}(\omega_+)$ breaking the SU(2)$\times$U(1) symmetry
of the ansatz. Locally, this term can be written as an exact form
\be
{\rm symmetry~breaking~term}:\quad d\L_{(2)}=d(\frac{8P}{r^2+P}\sqrt{1+\frac{\m}{P}}\o_{(2)})\,.
\ee
This means when solving the $B_{(2)}$ from the SU(2)$\times$U(1) invariant $\widetilde{H}_{(3)}$,
we need to supplement the ansatz with the symmetry breaking term -$\L_{(2)}$.
The thermodynamic quantities of the dyonic string in heterotic compactified on $T^4$
can be computed in the standard way outlined before. One only needs to be careful
when calculating the entropy. There is a Chern-Simons term hidden in $\widetilde{H}_{(3)}$ which will contribute
to the black string entropy. In order to evaluate its contribution, one should first recast the
Chern-Simons term in the form
\be
{\rm CS}_{(3)}(\omega_+)=\o^a_{\ b}(\omega_+)\wedge R^b_{\ a}(\omega_+)-\ft13\o^a_{\ b}(\omega_+)\wedge\o^b_{\ c}(\omega_+)\wedge\o^c_{\ a}(\omega_+)\,,
\ee
and then vary it with respect to the Riemann tensor when applying the Iyer-Wald formula.
The results are summarised below
\bea
M^{(\mathrm{het})}&=&\frac{ 3\pi  }{8}\mu+\frac{ \pi  }{4}(Q+P)-\frac{3 \pi  \mu ^2 (3 \mu +2 P) }{32 (\mu +P)^2 (\mu +2 P)}\alpha '\,,\cr
T^{(\mathrm{het})}&=&\frac{1}{2\pi}\frac{\sqrt{\mu }}{\sqrt{\mu +P} \sqrt{\mu +Q}}-\frac{\sqrt{\mu } P (5 \mu +4 P) }{4 \pi  \sqrt{\mu +Q} (\mu +P)^{5/2} (\mu +2 P)}\alpha '\,,\cr
S^{(\mathrm{het})}&=&\frac{1}{2} \pi ^2 \sqrt{\mu } \sqrt{\mu +P} \sqrt{\mu +Q}+\frac{\pi ^2 \sqrt{\mu }  \sqrt{\mu +Q}P  (5 \mu +4 P)}{4 (\mu +P)^{3/2} (\mu +2 P)}\alpha '\,,\cr
Q_{\rm e}^{(\mathrm{het})}&=&\frac{1}{4}\pi  \sqrt{Q} \sqrt{\mu +Q},\quad \Phi_{\rm e}^{(\mathrm{het})}=\sqrt{\frac{Q}{\mu +Q}}\,,\cr
Q_{\rm m}^{(\mathrm{het})}&=&\frac{ \pi}{4}  \sqrt{P} \sqrt{\mu +P},\quad \Phi_{\rm m}^{(\mathrm{het})}=\sqrt{\frac{P}{\mu +P}}+\frac{\mu  \sqrt{P}  (5 \mu +4 P)}{2 (\mu +P)^{5/2} (\mu +2 P)}\alpha '\,.
\eea
We can see that the thermodynamical quantities are related to those in the IIA case by interchanging parameters $P,\,Q$.
Thus obviously, they satisfy the first law of thermodynamics. Also the Euclidean action satifies
\be
I_E^{\rm het}=\beta G^{\rm het}\,,\quad G^{\rm het}=M^{(\mathrm{het})}-T^{(\mathrm{het})}S^{(\mathrm{het})}-\Phi_{\rm e}^{(\mathrm{het})}Q_{\rm e}^{(\mathrm{het})}\,.
\ee
We also obtained at fixed  conserved charges $(M,\,Q_{\rm e},\,Q_{\rm m})$
\be
\delta S^{\rm het}=\frac{\pi ^2 \sqrt{\mu }\sqrt{\mu +Q}(9 \mu +8P)}{16 (\mu +P)^{3/2}}\a'>0\,.
\ee

We now investigate the force felt by a probe fundamental string in the background of the dyonic string solutions with $\a'$ corrections.
The discussion is parallel to the one carried out in the IIA string setup. The purpose is that by comparison, one can
recognise some generic feature which may be shared by all consistent theory of quantum gravity.
Substituting the $\a'$-corrected dyonic string solution and expanding the potential in the region $r\gg {\rm max}\{\sqrt{\mu},\,\sqrt{Q},\,\sqrt{P}\}$
we find that the static potential takes the form
\be
\frac{{\cal V}(r)}{T_2}=1-\sqrt{1+\frac{\mu}{Q}}-\frac{\Sigma}{r^2}+{\cal O}(r^{-3})\,,\quad\Sigma=\frac{\mu }{2}+Q \left(1-\sqrt{1+\frac{\mu}{Q}}\right)\,.
\ee
In the Einstein frame, we
can separate the gravitational force from that of scalar exchange. The large $r$ expansions of the fields in Einstein frame
are given by
\bea
\sqrt{-g^E_{tt}g^E_{xx}}&=&1-\frac{\s_g}{ r^2}+{\cal O}(r^{-4})\,,\quad \s_g=\frac12\left(P+Q+\mu-\frac{3\a'(3 \mu+2 P)\mu ^2}{8(P+\m)^2(2P+\m)}\right)\,,\cr
L^{-\ft12}(r)&=&1-\frac{\s_L}{ r^2}+{\cal O}(r^{-4})\,,\quad\s_L=\frac12\left(Q-P+\frac{3\a'(3 \mu+2 P)\mu ^2 }{8(P+\m)^2(2P+\m)}\right)\,,
\cr
B_{tx}&=&\sqrt{1+\frac{\mu}{Q}}-\frac{\s_B}{r^{2}}+{\cal O}(r^{-4})\,,\quad \s_B=Q\sqrt{1+\frac{\mu}{Q}}\,,
\eea
thus $\Sigma=\s_g+\s_L-\s_B$.
Again parameters $Q,\,P$ are solved in terms of $Q_{\rm e},\,Q_{\rm m},\,\m$
\be
Q=\frac{\sqrt{\pi ^2 \mu ^2+64 Q_{\rm e}^2}-\pi  \mu }{2 \pi }\,,\quad P=\frac{\sqrt{\pi ^2 \mu ^2+64 Q_{\rm m}^2}-\pi  \mu }{2 \pi }\,.
\ee
Then there are three ways of solving $\m$ in terms of physical quantities.
\begin{description}
    \item[1)] We first express $\mu$ using $(T\,,Q_{\rm e}\,,Q_{\rm m})$ about $T=0$.  We find that
    \bea
    \mu&=&c_0T^2+c_2T^4+{\cal O}(T^6)\,,\cr
    c_0&=&64 Q_{\rm e} Q_{\rm m}+32 \pi  Q_{\rm e} \alpha'\,,\cr
    c_2&=&512\pi Q_{\rm e}Q_{\rm m}(Q_{\rm e}+Q_{\rm m})+128\pi^2Q_{\rm e}(4Q_{\rm m}+Q_{\rm e})\a'\,.
    \eea
Using the small $T$ expansion of $\m$, we obtain
\be
\Sigma=128 \pi  Q_{\rm e} Q_{\rm m}^2 T^4+128 \pi ^2 Q_{\rm e} Q_{\rm m} T^4 \a'\,.
\ee
This means the $\a'$ correction enforces the attractive force.
Dissembling $\Sigma$ into the contribution from gravity, dilaton and the 2-form, we have
\bea
   \s_g&=&\frac{2 (Q_{\rm e}+Q_{\rm m})}{\pi }+64 \pi Q_{\rm e}Q_{\rm m} \left(  Q_{\rm e}+ Q_{\rm m}\right) T^4
   +16\pi^2 Q_{\rm e}\left( Q_{\rm e}+4  Q_{\rm m}\right)T^4 \a'+{\cal O}(T^6)\,,\cr
   \s_L&=&\frac{2 (Q_{\rm e}-Q_{\rm m})}{\pi }-64\pi Q_{\rm e}Q_{\rm m} \left( Q_{\rm e}-Q_{\rm m}\right) T^4
   -16\pi^2 Q_{\rm e}\left(  Q_{\rm e}-4  Q_{\rm m}\right)T^4 \a'+{\cal O}(T^6)\,,\cr
   \s_B&=&\frac{4 Q_{\rm e}}{\pi}+{\cal O}(T^6)\,,
   \eea
We see that similar to the IIA case, at fixed electric and magnetic charges, the $\a'$ correction tends to increase the gravitational force when the system is slightly heated up from zero temperature. The sign of the $\a'$ correction to the scalar force depends on the relative magnitude between $Q_{\rm e}$ and $Q_{\rm m}$. However, their sum is always attractive.
 \item[2)]  In the second case, we express $\mu$ using $(M\,,Q_{\rm e}\,,Q_{\rm m})$ near the BPS limit in which $M$ approaches $M_0=Q_{\rm e}+Q_{\rm m}$.
    Thus we define the expansion parameter as $\delta M=M-M_0$. $\m$ can be expanded in terms of powers of $\delta M$
    \be
    \mu=c_1\delta m+c_2(\delta m)^2+c_3(\delta m)^3+{\cal O}(\delta m)^4\,,
    \ee
    where
    \be
    c_1=\frac{8}\pi\,,\quad c_2=\frac{-4 Q_{\rm m}^2-4 Q_{\rm e}Q_{\rm m}+3 \pi Q_{\rm e}\alpha' }{\pi Q_{\rm m}^2Q_{\rm e}}\,,\quad
    c_3=\frac{\pi}4c_2^2\,.
    \ee
    Meanwhile, the leading coefficient in the potential acquires an expansion
    \be
  \Sigma= \frac{2 \delta m^2}{\pi Q_{\rm e}}+ \left(\frac{3 }{2 'Q_{\rm e} Q_{\rm m}^2}\a'-\frac{2}{\pi Q_{\rm e}^2}-\frac{2}{\pi  Q_{\rm e} Q_{\rm m}}\right)\delta m^3
    \ee
    from which one can see that different from the IIA case, the higher derivative corrections appear to increase the attractive force but at the next leading order in the $\delta m$ expansion.
     In terms of the leading falloff coefficients of various fields in Einstein frame, up to first order in $\a'$
     we have
   \bea
   \s_g&=&\frac{2 (Q_{\rm e}+Q_{\rm m})}{\pi }+\frac{Q_{\rm e}+Q_{\rm m}}{Q_{\rm e}Q_{\rm m}}\frac{\delta m^2}{\pi}
   -\frac{3 \delta m^2 }{4 Q_{\rm m}^2}\a'\cr
   &&+\left(-\frac{1}{\pi  Q_{\rm e}^2}-\frac{1}{\pi  Q_{\rm m}^2}-\frac{2}{\pi Q_{\rm e} Q_{\rm m}}+\frac{3 \a' }{2 Q_{\rm e} Q_{\rm m}^2}+\frac{3 \a' }{2 Q_{\rm m}^3}\right)\delta m^3,,\cr
   \s_L&=&\frac{2 (Q_{\rm e}-Q_{\rm m})}{\pi }+\frac{Q_{\rm m}-Q_{\rm e}}{Q_{\rm e}Q_{\rm m}}\frac{\delta m^2}{\pi}
   +\frac{3 \delta m^2 }{4 Q_{\rm m}^2}\a'\cr
   &&+\left(-\frac{1}{\pi  Q_{\rm e}^2}-\frac{3 \a' }{2 Q_{\rm m}^3}+\frac{1}{\pi  Q_{\rm m}^2}\right)\delta m^3\cr
   \s_B&=&\frac{4 Q_{\rm e}}{\pi}+{\cal O}(\delta m^3)\,.
   \eea
   In this case we see that at fixed electric and magnetic charges, at order $\delta m^2$ the $\a'$ correction reduces the gravitational attractive force but increases the scalar attractive force. These two effects happen to cancel with each other. However, at the order $\delta m^3$, there is a nontrivial net contribution from $\a'$ correction which turns out to enforce the attractive force.
 \item[3)]  In the last case, we express $\mu$ using $(S\,,Q_{\rm e}\,,Q_{\rm m})$ near the BPS limit in which $S\rightarrow 0$.
    To distinguish with the entropy in the thermodynamics, we use $s$ to denote the small entropy expansion parameter. We find that
     $\mu$ can be expanded as
    \be
    \mu=c_0s^2+c_2s^4+{\cal O}(s^6)\,,
    \ee
    where up to first order in $\alpha$
    \be
    c_0=\frac{1}{4 \pi ^2 Q_{\rm e} Q_{\rm m}}-\frac{\alpha' }{8 \pi  Q_{\rm e} Q_{\rm m}^2}\,,\quad
     c_2= -\frac{Q_{\rm m}+Q_{\rm e}}{128 \pi ^3 Q_{\rm m}^3 Q_{\rm e}^3}+\frac{(4 Q_{\rm m}+7 Q_{\rm e})}{512 \pi ^2 Q_{\rm m}^4 Q_{\rm e}^3}\a'
    \ee
    Meanwhile, the leading coefficient in expansion of the potential acquires a small entropy expansion
    \be
   \Sigma=\frac{s^4}{512 \pi ^3 Q_{\rm e}^3 Q_{\rm m}^2}-\frac{s^4 }{512 \pi ^2 Q_{\rm e}^3 Q_{\rm m}^3}\alpha'+{\cal O}(s^6)
    \ee
    from which one can see that similar to the IIA case
    the higher derivative corrections appear to reduce the attractive force.
    In terms of the leading falloff coefficients of various fields in Einstein frame, up to first order in $\a'$
    we have
   \bea
   \s_g&=&\frac{2 (Q_{\rm e}+Q_{\rm m})}{\pi }+\frac{Q_{\rm e}+Q_{\rm m}}{1024 \pi ^3 Q_{\rm e}^3 Q_{\rm m}^3}s^4
   -\frac{4 Q_{\rm m}+7 Q_{\rm e}}{4096 \pi ^2 Q_{\rm m}^4 Q_{\rm e}^3}s^4\a'+{\cal O}(s^6)\,,\cr
   \s_L&=&\frac{2 (Q_{\rm e}-Q_{\rm m})}{\pi }+\frac{Q_{\rm m}-Q_{\rm e}}{1024 \pi ^3 Q_{\rm e}^3 Q_{\rm m}^3}s^4
   +\frac{7 Q_{\rm e}-4 Q_{\rm m}}{4096 \pi ^2 Q_{\rm m}^4 Q_{\rm e}^3}s^4\a'+{\cal O}(s^6)\,,\cr
   \s_B&=&\frac{4 Q_{\rm e}}{\pi}+{\cal O}(s^6)\,.
   \eea
   In this case we see that at fixed electric and magnetic charges, the $\a'$ correction reduces the gravitational force
  when the entropy of the system is slightly above 0. The sign of the $\a'$ correction to the scalar force depends on the relative magnitude between $Q_{\rm e}$ and $Q_{\rm m}$. However, their sum is always repulsive.

\end{description}

In summary, we have studied the static potential in the probe string action caused by the macroscopic dyonic solutions
in both IIA on K3 and heterotic on 4-torus. We noticed that at fixed electric and magnetic charges, if the inclusion of
the $\a'$ correction is an isoentropic process, the higher derivative corrections always tend to reduce the attractive force
when the system is slight away from extremality. On the other hand, if the inclusion of
the $\a'$ correction is an isothermoal process, the higher derivative corrections always enforce the gravitational contribution to the attractive force.

\section{Conclusion }

In many examples of supergravity models, the higher derivative corrections do not affect the charge-to-mass ratio of extremal (BPS) black holes. Thus apparently previous theoretical constraints on the higher derivative couplings implied by the WGC do not seem to apply.
In order to provide a new criteria for which kind of higher derivative couplings in supergravities is consistent with quantum gravity, we studied non-extremal dyonic strings in two string theory setups where the leading higher derivative corrections are known. One of them is type IIA string compactified on K3, the other is heterotic string compactified on 4-torus. These two 6D string theories are related to each other by a $S$-duality. In other words, the fundamental string in one theory becomes the solitonic string in the other after the duality transformation. In these two setups, we constructed non-extremal dyonic string solutions $\a'$-corrections. We computed the thermodynamical quantities associated with the non-extremal string and verified that they satisfied the first law of thermodynamics. Euclidean actions of the solutions were also calculated.

We tried to answer the previous question by studying the force felt by a probe fundamental string in the background of the macroscopic non-extremal dyonic strings. We found that away from extremality, the attractive force overwhelms the repulsive force. However, near extremality, the higher derivative corrections can reduce the difference between the attractive force and repulsive force, if the inclusion of higher derivative correction is viewed as an isoentropic process.

We also noticed that if the inclusion of higher derivative correction is viewed as an isothermal process, the higher derivative corrections tend to enforce the attractive force in both scenarios. At the moment, we do not fully understand the reason behind these phenomenon. Our results contradict with the standard isothermal process in the WGC discussion of Einstein-Maxwell theory where one calculates the $\delta M$ and hence the change of force by requiring the black hole to be extremal (zero temperature) before and after the higher-order corrections. It should be interesting to explore further in the future in other setups with string theory embedding and see if similar features appear. After all, certain higher derivative extensions of supergravity models with string theory origin are also known in $D=4,\,5$ \cite{Bobev:2021oku, Ozkan:2013nwa}.

\section*{Acknowledgement}

We are grateful to Stefan Theisen for useful discussion. The work is supported in part by the National Natural Science Foundation of China (NSFC) grants No.~11875200, No.~11935009 and No.~12175164.

\appendix

\section{Perturbative black dyonic strings}

In this section, we present the dynonic string solutions to the perturbative field equations up to first order in $\a'$ in both the type IIA and heterotic setups.

\subsection{Type IIA case }
In the type IIA case, we present the full solutions to the perturbations without imposing any boundary conditions. The solutions involve eight integration constants $C_i$, $i=1,2,\ldots, 8)$. We first define
\bea
C_0&=&2 C_1 Q+C_3 Q (\mu +2 Q)-2 \left(C_4+C_7\right) Q (\mu +Q)+2 C_5 (\mu +Q)\,,\nn\\
X(r)&=&(1+\frac{Q}{\mu })^{-2}\left((3+\frac{2Q}{\mu })\log (1-\frac{\mu }{r^2})-\frac{\mu^2}{Q^2}\log (1+\frac{Q}{r^2})\right)\,.
\eea
Solutions to various perturbations are given by
\bea
&&\delta\omega=\frac{\mu  \sqrt{P} \sqrt{\mu +P} }{2 r^2 (P+r^2) (Q+r^2)}\a'\,,\nn\w2
\label{equ coupling f 8}
&&\delta f=\frac{C_1 }{r^2}(1+\frac{P}{r^2})-\frac{\mu  \a'}{8 Q r^4}(1+\frac{Q}{\mu })^{-1}\Big(\mu +2 Q-\frac{Q (5 \mu +4 Q)-P (\mu +2 Q)}{r^2}\cr
&&-\frac{Q (\mu  P+4 Q^2+4 \mu  Q)}{r^4}\Big)(1+\frac{Q}{r^2})^{-2}+\frac{C_2}{2 \mu  r^2}(1+\frac{P}{r^2})\log (1-\frac{\mu }{r^2})-\frac{\a'}{8r^2}(1+\frac{P}{r^2})X(r),\nn\w2
&&\delta h=\frac{C_1}{\mu  r^2}(\mu +2 P-\frac{\mu  P}{r^2})+\frac{C_2}{\mu  r^2}(1+\frac{P}{r^2})+C_7 (1-\frac{\mu }{r^2})-\frac{\a'}{8 Q r^2}(1+\frac{Q}{\mu })^{-1}\Big(2 (\mu +2 Q)\cr
&&+\frac{-\mu ^2+4 P (\mu +2 Q)+4 Q^2+2 \mu  Q}{r^2}+\frac{2 P^2 (\mu +2 Q)-2 \mu  P (\mu +2 Q)-4 \mu  Q (\mu +Q)}{r^4}\cr
&&+\frac{P }{r^6}\Big(P (-\mu ^2+4 Q^2+2 \mu  Q)+4 \mu  Q (\mu +Q)\Big)\Big)(1+\frac{Q}{r^2})^{-1}(1+\frac{P}{r^2})^{-1}\cr
&&+\frac{1}{2 \mu ^3}(1+\frac{Q}{\mu })^{-1}\Big( (C_0 \mu +4 C_1 P Q)(1-\frac{\mu }{r^2})-C_2\Big(\mu -2 Q-\frac{2 P (\mu +Q)+\mu  (2 \mu -Q)}{r^2}\cr
&&+\frac{\mu  P (\mu +Q)}{r^4}\Big)
\Big)\log (1-\frac{\mu }{r^2})-\frac{ \a'}{8 \mu }(2-\frac{\mu }{r^2})(1+\frac{P}{r^2})X(r),\nn\w2
&&\delta D=-\Big(\frac{C_0}{2 \mu  r^2}+\frac{ C_1 P Q}{\mu ^2 r^2}(2+\frac{\mu }{r^2})+\frac{C_2 Q}{\mu ^2 r^2}(2-\frac{P}{r^2})-\frac{C_3 Q}{2 r^2}-\frac{C_4}{2}(1+\frac{Q}{r^2})\Big)(1+\frac{Q}{r^2})^{-2}\cr
&&+\frac{\a'}{8 \mu  Q r^2}(1+\frac{Q}{\mu })^{-1}\Big((\mu +2 Q) (3 \mu +4 Q)+\frac{4 Q^2}{r^2} \Big((P+4 Q)+\mu ^2 (3 P+7 Q)\cr
&&+4 \mu  Q (2 P+5 Q)\Big) +\frac{2 Q}{r^4} \Big(2 P (\mu +Q) (\mu +2 Q)-\mu  Q (2 \mu +3 Q)-P^2 (\mu +2 Q)\Big)\cr
&&+\frac{P Q }{r^6}\Big(\mu ^2 (P-11 Q)-4 \mu  Q (P+4 Q)-4 Q^2 (2 P+Q)\Big)\cr
&&-\frac{P Q^2 }{r^8}\Big(P (4 Q^2+2 \mu  Q-\mu ^2)+4 \mu  Q (\mu +Q)\Big)
\Big)(1+\frac{Q}{r^2})^{-4}(1+\frac{P}{r^2})^{-1}\cr
&&-\frac{1}{2 \mu ^3}\Big(C_0 \mu +4 C_1 P Q+C_2 Q(2-\frac{\mu }{r^2})(2-\frac{P}{r^2}
\Big)(1+\frac{Q}{r^2})^{-2}\log (1-\frac{\mu }{r^2})\cr
&&+\frac{\a'}{8 \mu^2}\Big(3 \mu +4 Q-\frac{2 Q (\mu +P)}{r^2}+\frac{\mu  P Q}{r^4}\Big)(1+\frac{Q}{r^2})^{-2}X(r),\nn\w2
&&\delta A=C_6-\Big(\frac{C_0 Q}{2 \mu ^2 r^2}(1+\frac{Q}{\mu })^{-1}(1-\frac{\mu }{r^2})
+\frac{C_1 P Q}{\mu ^3 r^2}(1+\frac{Q}{\mu })^{-1}(2Q+\frac{\mu  (\mu -Q)}{r^2})\cr
&&+\frac{C_2 Q}{2 \mu ^3 r^2}(1+\frac{Q}{\mu })^{-1}(\mu+4 Q -\frac{2 P (\mu +Q)+3 \mu  Q}{r^2})+\frac{C_3 Q^2}{2 r^4}+\frac{C_5}{2 r^2}(1+\frac{Q}{r^2})\Big)(1+\frac{Q}{r^2})^{-2}\cr
&&+\frac{\a'}{8 \mu  r^2}(1+\frac{Q}{\mu })^{-1}\Big(4 (\mu +2 Q)-\frac{5 \mu ^2+2 P (\mu +2 Q)-16 Q^2-2 \mu  Q}{r^2}\cr
&&-\frac{1}{r^4}\Big(P (-\mu ^2+8 Q^2+4 \mu  Q)+5 \mu  Q (3 \mu +4 Q)
\Big) -\frac{Q }{r^6}\Big(P (-\mu ^2+4 Q^2+2 \mu  Q)\cr
&&+2 \mu  Q (\mu +Q)\Big) \Big)(1+\frac{Q}{r^2})^{-4}-\frac{Q}{8 \mu ^4}(1+\frac{Q}{\mu })^{-1}\Big(4 C_0 \mu  (1-\frac{\mu }{r^2})+16 C_1 P Q (1-\frac{\mu }{r^2})\cr
&&+4 C_2\Big(\mu +4 Q-\frac{2 P (\mu +Q)+\mu  (2 \mu +5 Q)}{r^2}+\frac{\mu  P (\mu +Q)}{r^4}
\Big)\Big)(1+\frac{Q}{r^2})^{-2}\log(1-\frac{\mu }{r^2})\cr
&&+\frac{Q\a'}{8 \mu^2}(4-\frac{5 \mu +2 P}{r^2}+\frac{\mu  P}{r^4})(1+\frac{Q}{r^2})^{-2}X(r),\nn\w2
\label{equ coupling L 8}
&&\delta L=\Big(\frac{C_0 \mu +2 C_1 P (2 Q-\mu )+2 C_2 (P+2 Q)-C_3 \mu ^2Q}{2 \mu ^2 r^2}
+C_8(1+\frac{Q}{r^2})\Big)(1+\frac{P}{r^2})^{-1}\cr
&&-\frac{\a'}{8 \mu  Q r^2}(1+\frac{Q}{\mu })^{-1}\Bigg((\mu +2 Q) (3 \mu +2 P+4 Q)
-\frac{1}{r^2}\Big(-8 Q^2 (P+2 Q)+\mu ^2 (P-7 Q)\cr
&&-4 \mu  Q (P+5 Q)\Big)-\frac{Q }{r^4}\Big(P (\mu ^2-4 Q^2-2 \mu  Q)+2 \mu  Q (2 \mu +3 Q)\Big)\Bigg)(1+\frac{Q}{r^2})^{-2}(1+\frac{P}{r^2})^{-1}\cr
&&+\frac{1}{8 \mu ^4}(1+\frac{Q}{\mu })^{-1}\Bigg(
2 (C_0 \mu +4 C_1 P Q)(\mu +2 Q-\frac{\mu  Q}{r^2})+2 C_2\Big(
3 \mu ^2+4 P (\mu +Q)+8 Q^2\cr
&&+8 \mu  Q-\mu\frac{ 2 P (\mu +Q)+Q (\mu +4 Q) }{r^2}\Big)\Bigg)(1+\frac{P}{r^2})^{-1}\log (1-\frac{\mu }{r^2})\cr
&&-\frac{\a'}{8 \mu^2}\Big(
3 \mu +2 P+4 Q-\frac{\mu  (P+2 Q)}{r^2}\Big)(1+\frac{P}{r^2})^{-1}X(r).
\eea

\subsection{Heterotic case}

This solution is obtained by performing the $S$-duality of the type IIA solution, after imposing the appropriate boundary condition on the horizon and at the asymptotic infinity:
\bea
&&\delta\omega=-\frac{\mu  \sqrt{Q}  \sqrt{\mu +Q}}{2 r^2 (P+r^2) (Q+r^2)}\alpha'\,,\nn\w2
&&\delta f=\frac{\alpha '}{4 P (\mu +P)^2 (\mu +2 P) (P+r^2)^3}\Big(3 \mu ^2 P  (3 \mu +2 P)r^4+\mu ^2  (\mu ^3+44 \mu  P^2\cr
&&+28 P^3+5 \mu ^2 P)r^2 -3 P^2 (7 \mu ^4+58 \mu ^2 P^2+56 \mu  P^3+16 P^4+19 \mu ^3 P)+\frac{P^2 }{r^2}(17 \mu ^5\cr
&&+92 \mu ^2 P^3+170 \mu ^3 P^2 -16 \mu  P^4-16 P^5+83 \mu ^4 P)+\frac{\mu  P^3}{r^4} (2 \mu +P) (\mu ^3+16 \mu  P^2\cr
&&+8 P^3+6 \mu ^2 P)\Big)-\frac{\mu ^4 \alpha '}{4 P^2 r^2 (\mu +P)^2}(1+\frac{P}{r^2}) \log (1+\frac{P}{r^2})\,,\nn \w2
&&\delta h=\frac{\mu  \alpha '}{4 P (\mu +P)^2 (\mu +2 P) (P+r^2)^2 (Q+r^2)}\Big(4 \mu ^2 P^2 (6 P+5 Q)+4 \mu  P^3 (P+5 Q)\cr
&&+8 P^4 Q +\mu ^3 P (29 P+13 Q)+2 \mu ^4 (3 P+Q)+\mu ^2 (2 \mu ^2+4 P^2+9 \mu  P-2 P Q-\mu  Q) r^2\cr
&&-\mu ^2  (\mu +2 P)r^4 +\frac{P}{r^2} \Big(\mu ^2 P^2 (20 Q-\mu )+8 P^4 (Q-3 \mu )+2 \mu  P^3 (8 Q-9 \mu )-8 P^5\cr
&&+17 \mu ^3 P Q+2 \mu ^4 Q\Big)-\frac{P^2 Q}{r^4} (4 \mu ^4+38 \mu ^2 P^2+32 \mu  P^3+8 P^4+17 \mu ^3 P)\Big)
\Big)\cr
&&+\frac{\mu ^3 \alpha '}{4 P^2 (\mu +P)^2}(1-\frac{2 (\mu +P)}{r^2}+\frac{\mu  P}{r^4}) \log (1+\frac{P}{r^2})\,,\nn\w2
&&\delta D=-\frac{Q \alpha '}{4 P (\mu +P)^2 (\mu +2 P) (P+r^2)^2 (Q+r^2)^2}\Big(P^2 (4 \mu ^4+38 \mu ^2 P^2+32 \mu  P^3\cr
&&+8 P^4+17 \mu ^3 P)+\mu ^2  (\mu +2 P)r^6-\mu ^2  (\mu +4 P) (2 \mu +P)r^4+\mu  P (4 P^3-2 \mu ^3\cr
&&-4 \mu  P^2-13 \mu ^2 P) r^2\Big)+\frac{\mu ^2 Q r^4 \alpha '}{4 P^2 (\mu +P)^2 (Q+r^2)^2}(1-\frac{2 (\mu +P)}{r^2}+\frac{\mu  P}{r^4}) \log (1+\frac{P}{r^2})\,,\nn\w2
&&\delta A=\delta D+\frac{\mu  Q }{2 (P+r^2) (Q+r^2)^2}\alpha'\,,\nn\w2
&&\delta L=\frac{\alpha '}{8 P (\mu +P)^2 (\mu +2 P) (P+r^2)^3}\Big(\mu ^2 (3 \mu ^2+14 P^2+19 \mu  P+4 P Q+2 \mu  Q)r^4\cr
&&+\mu ^2  (6 P^2 (5 P+Q)+\mu ^2 (7 P-Q)+\mu  P (41 P+Q))r^2\cr
&&-P (56 \mu  P^4+2 \mu ^2 P^2 (28 P-Q)+16 P^5+\mu ^3 P (18 P+Q)+\mu ^4 (4 P+Q))
\Big)\cr
&&+\frac{\mu ^2 \alpha '}{8 P^2 (\mu +P)^2 (P+r^2)} (2 \mu  P-4 P r^2+\mu  Q-2 Q r^2-3 \mu  r^2)\log(1+\frac{P}{r^2})\,.
\eea

\section{Another solution}

As was explained in section \ref{sec:2aonk3},  there were two independent four-derivative supersymmetric extensions to 6D supergravity. The general bosonic Lagrangian of supergravity is therefore given by
\be
{\cal L}={\cal L}_0 + \fft1{16} \Big(\lambda_{\rm GB} {\cal L}_{\rm GB} +\lambda _{\text{Riem}^2} {\cal L}_{{\rm Riem}^2}\Big).
\ee
where ${\cal L}_{\rm GB}$ and ${\cal L}_{{\rm Riem}^2}$ are given in \eqref{4derivaction}. Type IIA on K3 is given by $\lambda_{\rm GB}=\lambda _{\text{Riem}^2}=\alpha'$, which is dual to heterotic string on $T^4$. The most general linearized static dyonic string solution would involve five parameters $(\mu, Q,P,\lambda_{\rm GB}, \lambda_{{\rm {Riem}^2}}$), which we do not find analytic solutions. When the two couplings are equal, we obtain the general dyonic string solution which is the focus of our discussion in the main text.  Here we present another special case with two independent couplings but equal charges $Q=P=Q_P$. The solution involves eight integration constants $C_i$, $i=1,2,\ldots, 8$ and for convenience we define
\bea
&&C_0=2 C_1 Q_P+C_3 Q_P \left(\mu +2 Q_P\right)-2 \left(C_4+C_7\right) Q_P \left(\mu +Q_P\right)+2 C_5 \left(\mu +Q_P\right)\,,\nn\\
&&X(r)=(1+\frac{Q_P}{\mu })^{-2}\left((3+\frac{2Q_P}{\mu })\log (1-\frac{\mu }{r^2})-\frac{\mu^2}{Q_P^2}\log (1+\frac{Q_P}{r^2})\right)\,.
\eea
Solution to the linear order of the couplings is given by
\bea
\label{equ charge f 8}
&&\delta\omega=\frac{\mu  \sqrt{P} \sqrt{\mu +P} }{4 r^2 (P+r^2) (Q+r^2)}(\lambda _{\text{GB}}+\lambda _{\text{Riem}^2})\,,\nn\w2
&&\delta f=\frac{C_1 }{r^2}(1+\frac{Q_P}{r^2})-\frac{\mu (\lambda _{\text{GB}}+\lambda _{\text{Riem}^2})}{16 r^4 Q_P}(1+\frac{Q_P}{\mu })^{-1}(1+\frac{Q_P}{r^2})^{-1} \Big(\mu+2 Q_P \cr
&&-\frac{Q_P (5 \mu +4 Q_P)}{r^2}\Big)+\frac{C_2}{2 \mu  r^2}\left(1+\frac{Q_P}{r^2}\right)\log \left(1-\frac{\mu }{r^2}\right)-\frac{\left(\lambda _{\text{GB}}+\lambda _{\text{Riem}^2}\right)}{16 r^2}\left(1+\frac{Q_P}{r^2}\right)X(r)\,,\nn\w2
&&\delta h=\frac{C_1}{\mu  r^2}(\mu +2 Q_P-\frac{\mu  Q_P}{r^2})+\frac{C_2}{\mu  r^2}(1+\frac{Q_P}{r^2})+C_7(1-\frac{\mu }{r^2})\cr
&&-\frac{\lambda _{\text{GB}}+\lambda _{\text{Riem}^2}}{16 r^2 Q_P}(1+\frac{Q_P}{\mu })^{-1}\Big(
2 (\mu +2 Q_P)+\frac{-\mu ^2+6 \mu  Q_P+12 Q_P^2}{r^2}\cr
&&+\frac{2 Q_P (-3 \mu ^2-3 \mu  Q_P+2 Q_P^2)}{r^4}+\frac{Q_P^2 (3 \mu ^2+6 \mu  Q_P+4 Q_P^2)}{r^6}\Big)(1+\frac{Q_P}{r^2})^{-2}\cr
&& +\frac{1}{2 \mu ^3}(1+\frac{Q_P}{\mu })^{-1}\Big( (C_0 \mu +4 C_1 Q_P^2)(1-\frac{\mu }{r^2})\cr
&&-C_2(\mu -2 Q_P-\frac{2 \mu ^2+\mu  Q_P+2 Q_P^2}{r^2}+\frac{\mu  Q_P (\mu +Q_P)}{r^4})\Big)\log (1-\frac{\mu }{r^2})\cr
&&-\frac{ (\lambda _{\text{GB}}+\lambda _{\text{Riem}^2})}{16\mu }(2-\frac{\mu }{r^2})(1+\frac{Q_P}{r^2})X(r)\,,\nn\w2
&&\delta D=-\frac{1}{2}\Big(\frac{C_0}{\mu  r^2}+\frac{2 (C_1\mu -C_2) Q_P^2}{\mu ^2 r^4}+\frac{4 Q_P (C_1 Q_P+C_2)}{\mu ^2 r^2}-\frac{C_3 Q_P}{r^2}\cr
&&-C_4 (1+\frac{Q_P}{r^2})\Big)(1+\frac{Q_P}{r^2})^{-2} +\frac{\lambda _{\text{GB}}+\lambda _{\text{Riem}^2}}{16 r^2 (\mu +Q_P)}\Big(10 \mu +\frac{3 \mu ^2}{Q_P}+8 Q_P\cr
&&+\frac{7 \mu ^2+18 \mu  Q_P+12 Q_P^2}{r^2}-\frac{Q_P (7 \mu ^2+14 \mu  Q_P+8 Q_P^2)}{r^4}\cr
&&-\frac{Q_P^2 (3 \mu ^2+6 \mu  Q_P+4 Q_P^2)}{r^6}\Big)(1+\frac{Q_P}{r^2})^{-4}-\frac{1}{2 \mu ^3}\Big( C_0 \mu +4 C_1 Q_P^2+C_2 (4 Q_P\cr
&&-\frac{2 Q_P (\mu +Q_P)}{r^2}+\frac{\mu  Q_P^2}{r^4}) \Big)(1+\frac{Q_P}{r^2})^{-2}\log(1-\frac{\mu }{r^2})\cr
&&+\frac{\lambda _{\text{GB}}+\lambda _{\text{Riem}^2}}{16 \mu^2}\Big(3 \mu +4 Q_P-\frac{2 Q_P (\mu +Q_P)}{r^2}+\frac{\mu  Q_P^2}{r^4}\Big)(1+\frac{Q_P}{r^2})^{-2}X(r)\,,\nn\w2
&&\delta A=C_6-\frac{C_5}{2 \mu ^2 r^2}(\mu ^2+2 \mu  Q_P+2 Q_P^2+\frac{2 Q_P^3-\mu ^2 Q_P}{r^2})(1+\frac{Q_P}{r^2})^{-2}\cr
&&-\frac{Q_P}{2 \mu ^3 r^2}(1+\frac{Q_P}{\mu })^{-1}\Big(C_0 \mu  (1-\frac{\mu }{r^2})+2 C_1 Q_P (2 Q_P+\frac{\mu  (\mu -Q_P)}{r^2})+C_2 (\mu+4 Q_P\cr
&&-\frac{Q_P (5 \mu +2 Q_P)}{r^2})+\frac{C_3 \mu ^2 Q_P (\mu +Q_P)}{r^2}-\frac{2 C_5 (\mu +Q_P){}^2 (r^2-\mu +Q_P)}{r^2}\Big)(1+\frac{Q_P}{r^2})^{-2}\cr
&&+\frac{\lambda _{\text{GB}}+\lambda _{\text{Riem}^2}}{16 \mu  r^2}(1+\frac{Q_P}{\mu })^{-1}\Big(
4 (\mu +2 Q_P)+\frac{12 Q_P^2-5 \mu ^2}{r^2}\cr
&&-\frac{2 Q_P (7 \mu ^2+12 \mu  Q_P+4 Q_P^2)}{r^4}-\frac{Q_P^2 (\mu +2 Q_P)^2}{r^6}\Big)(1+\frac{Q_P}{r^2})^{-4}\cr
&&-\frac{Q_P}{2 \mu ^4}(1+\frac{Q_P}{\mu })^{-1}\Big( (C_0 \mu +4 C_1 Q_P^2)(1-\frac{\mu }{r^2})\cr
&&+C_2(\mu+4 Q_P -\frac{2 \mu ^2+7 \mu  Q_P+2 Q_P^2}{r^2}+\frac{\mu  Q_P (\mu +Q_P)}{r^4})
 \Big)(1+\frac{Q_P}{r^2})^{-2}\log (1-\frac{\mu }{r^2})\cr
&&+\frac{\lambda _{\text{GB}}+\lambda _{\text{Riem}^2}}{16 \mu^2}(
 4-\frac{5 \mu +2 Q_P}{r^2}+\frac{\mu  Q_P}{r^4})(1+\frac{Q_P}{r^2})^{-2}X(r)\,,\nn\w2
\label{equ charge L 8}
&&\delta L=C_8+\frac{C_0 \mu -(C_3 \mu ^2+2 C_1 \mu -6 C_2) Q_P+4 C_1 Q_P^2}{2 \mu ^2 r^2}(1+\frac{Q_P}{r^2})^{-1}\cr
&&-\frac{\lambda _{\text{GB}}+\lambda _{\text{Riem}^2}}{16 \mu  r^2 Q_P}(1+\frac{Q_P}{\mu })^{-1} \Big(3 (\mu +2 Q_P){}^2+\frac{6 Q_P (\mu +2 Q_P){}^2}{r^2}\cr
&&+\frac{Q_P^2 (-5 \mu ^2-4 \mu  Q_P+4 Q_P^2)}{r^4}
\Big)(1+\frac{Q_P}{r^2})^{-3}+\frac{C_0 \mu +3 C_2 (\mu +2 Q_P)+4 C_1 Q_P^2}{4 \mu ^4}\cr
&&\times (1+\frac{Q_P}{\mu })^{-1}(\mu +2 Q_P-\frac{\mu  Q_P}{r^2}
)(1+\frac{Q_P}{r^2})^{-1}\log (1-\frac{\mu }{r^2})\cr
&&-\frac{3 (\lambda _{\text{GB}}+\lambda _{\text{Riem}^2})}{16 \mu^2}(\mu +2 Q_P-\frac{\mu  Q_P}{r^2}
)(1+\frac{Q_P}{r^2})^{-1}X(r)\,.
\eea
Requiring the solution be regular on the horizon and asymptotic to Minkowski fixes the parameters to be
\bea
\label{solution 2}
&&\{C_1,\,C_2,\,C_3,\,C_4,\,C_5,\,C_6,\,C_7,\,C_8\}\cr
&=&\{0,\frac{\mu ^2 (3 \mu +2 Q_P) \left(\lambda _{\text{GB}}+\lambda _{\text{Riem}^2}\right)}{8 (\mu +Q_P)^2},\frac{3 \mu ^2 (3 \mu +2 Q_P)\left(\lambda _{\text{GB}}+\lambda _{\text{Riem}^2}\right) }{8 Q_P (\mu +Q_P)^2 (\mu +2 Q_P)}, 0,0,0,0,0\}\,.
\eea
It should be noticed that when $\lambda _{\text{GB}}+\lambda _{\text{Riem}^2}=0$, the expressions above all vanish, the solution therefore receives no correction.


\begin{thebibliography}{99}
\bibitem{Arkani-Hamed:2006emk}
N.~Arkani-Hamed, L.~Motl, A.~Nicolis and C.~Vafa,
``The string landscape, black holes and gravity as the weakest force,''
JHEP \textbf{06} (2007), 060, arXiv:hep-th/0601001 [hep-th].

\bibitem{Polchinski:2003bq}
J.~Polchinski,
``Monopoles, duality, and string theory,''
Int. J. Mod. Phys. A \textbf{19S1} (2004), 145-156,
arXiv:hep-th/0304042 [hep-th].

\bibitem{Banks:2010zn}
T.~Banks and N.~Seiberg,
``Symmetries and Strings in Field Theory and Gravity,''
Phys. Rev. D \textbf{83} (2011), 084019, arXiv:1011.5120 [hep-th].

\bibitem{Garg:2018reu}
S.~K.~Garg and C.~Krishnan,
``Bounds on Slow Roll and the de Sitter Swampland,''
JHEP \textbf{11} (2019), 075, arXiv:1807.05193 [hep-th].

\bibitem{Ooguri:2018wrx}
H.~Ooguri, E.~Palti, G.~Shiu and C.~Vafa,
``Distance and de Sitter Conjectures on the Swampland,''
Phys. Lett. B \textbf{788} (2019), 180-184, arXiv:1810.05506 [hep-th].

\bibitem{Harlow:2018tng}
D.~Harlow and H.~Ooguri,
``Symmetries in quantum field theory and quantum gravity,''
Commun. Math. Phys. \textbf{383} (2021) no.3, 1669-1804, arXiv:1810.05338 [hep-th].

\bibitem{Palti:2019pca}
E.~Palti,
``The swampland: introduction and review,''
Fortsch. Phys. \textbf{67} (2019) no.6, 1900037,arXiv:1903.06239 [hep-th].

\bibitem{Kats:2006xp}
Y.~Kats, L.~Motl and M.~Padi,
``Higher-order corrections to mass-charge relation of extremal black holes,''
JHEP \textbf{12} (2007), 068, arXiv:hep-th/0606100 [hep-th].


\bibitem{Cheung:2018cwt}
C.~Cheung, J.~Liu and G.N.~Remmen,
``Proof of the weak gravity conjecture from black hole entropy,''
JHEP \textbf{10} (2018), 004,
arXiv:1801.08546 [hep-th].

\bibitem{Hamada:2018dde}
Y.~Hamada, T.~Noumi and G.~Shiu,
``Weak Gravity Conjecture from Unitarity and Causality,''
Phys. Rev. Lett. \textbf{123} (2019) no.5, 051601,arXiv:1810.03637 [hep-th].

\bibitem{Cheung:2019cwi}
C.~Cheung, J.~Liu and G.N.~Remmen,
``Entropy bounds on effective field theory from rotating dyonic black holes,''
Phys. Rev. D \textbf{100} (2019) no.4, 046003, arXiv:1903.09156 [hep-th].

\bibitem{Charles:2019qqt}
A.M.~Charles,
``The weak gravity conjecture, RG flows, and supersymmetry,'' arXiv:1906.07734 [hep-th].

\bibitem{Loges:2019jzs}
G.J.~Loges, T.~Noumi and G.~Shiu,
``Thermodynamics of 4D dilatonic black holes and the weak gravity conjecture,''
Phys. Rev. D \textbf{102} (2020) no.4, 046010
arXiv:1909.01352 [hep-th].

\bibitem{Cremonini:2019wdk}
S.~Cremonini, C.R.T.~Jones, J.T.~Liu and B.~McPeak,
``Higher-derivative corrections to entropy and the weak gravity conjecture in anti-de Sitter space,''
JHEP \textbf{09} (2020), 003, arXiv:1912.11161 [hep-th].

\bibitem{Ma:2020xwi}
L.~Ma, Y.Z.~Li and H.~L\"u,
``$D = 5$ rotating black holes in Einstein-Gauss-Bonnet gravity: mass and angular momentum in extremality,''
JHEP \textbf{01}, 201 (2021), arXiv:2009.00015 [hep-th].



\bibitem{Cano:2019oma}
P.A.~Cano, T.~Ort\'\i{}n and P.F.~Ramirez,
``On the extremality bound of stringy black holes,''
JHEP \textbf{02} (2020), 175, arXiv:1909.08530 [hep-th].

\bibitem{Cano:2019ycn}
P.A.~Cano, S.~Chimento, R.~Linares, T.~Ort\'\i{}n and P.F.~Ram\'\i{}rez,
``$\alpha'$ corrections of Reissner-Nordstr\"om black holes,''
JHEP \textbf{02} (2020), 031, arXiv:1910.14324 [hep-th].

\bibitem{Liu:2013dna}
J.T.~Liu and R.~Minasian,
``Higher-derivative couplings in string theory: dualities and the $B$-field,''
Nucl. Phys. B \textbf{874} (2013), 413-470, arXiv:1304.3137 [hep-th].


\bibitem{Bergshoeff:1985mz}
  E.~Bergshoeff, E.~Sezgin and A.~Van Proeyen,
  ``Superconformal tensor calculus and matter couplings in six-dimensions,''
  Nucl.\ Phys.\ B {\bf 264} (1986) 653, Erratum: [Nucl.\ Phys.\ B {\bf 598} (2001) 667].

\bibitem{Butter:2016qkx}
D.~Butter, S.M.~Kuzenko, J.~Novak and S.~Theisen,
``Invariants for minimal conformal supergravity in six dimensions,''
JHEP \textbf{12} (2016), 072, arXiv:1606.02921 [hep-th].

\bibitem{Butter:2017jqu}
D.~Butter, J.~Novak and G.~Tartaglino-Mazzucchelli,
``The component structure of conformal supergravity invariants in six dimensions,''
JHEP \textbf{05} (2017), 133, arXiv:1701.08163 [hep-th].

\bibitem{Novak:2017wqc}
J.~Novak, M.~Ozkan, Y.~Pang and G.~Tartaglino-Mazzucchelli,
``Gauss-Bonnet supergravity in six dimensions,''
Phys. Rev. Lett. \textbf{119} (2017) no.11, 111602, arXiv:1706.09330 [hep-th].

\bibitem{Butter:2018wss}
D.~Butter, J.~Novak, M.~Ozkan, Y.~Pang and G.~Tartaglino-Mazzucchelli,
``Curvature squared invariants in six-dimensional ${\cal N} = (1,0)$ supergravity,''
JHEP \textbf{04} (2019), 013, arXiv:1808.00459 [hep-th].

\bibitem{Heidenreich:2019zkl}
B.~Heidenreich, M.~Reece and T.~Rudelius,
``Repulsive Forces and the Weak Gravity Conjecture,''
JHEP \textbf{10} (2019), 055, arXiv:1906.02206 [hep-th].

\bibitem{Pang:2019qwq}
Y.~Pang,
``Attractor mechanism and nonrenormalization theorem in 6D (1, 0) supergravity,''
Phys. Rev. D \textbf{103} (2021) no.2, 026018, arXiv:1910.10192 [hep-th].

\bibitem{Bergshoeff:2012ax}
E.~Bergshoeff, F.~Coomans, E.~Sezgin and A.~Van Proeyen,
``Higher derivative extension of 6D chiral gauged supergravity,''
JHEP \textbf{07} (2012), 011, arXiv:1203.2975 [hep-th].

\bibitem{Iyer:1994ys}
V.~Iyer and R.M.~Wald,
``Some properties of Noether charge and a proposal for dynamical black hole entropy,''
Phys. Rev. D \textbf{50} (1994), 846-864,
arXiv:gr-qc/9403028 [gr-qc].

\bibitem{Brown:1992br}
J.D.~Brown and J.W.~York, Jr.,
``Quasilocal energy and conserved charges derived from the gravitational action,''
Phys. Rev. D \textbf{47} (1993), 1407-1419, arXiv:gr-qc/9209012 [gr-qc].

\bibitem{Duff:1994an}
M.J.~Duff, R.R.~Khuri and J.X.~Lu,
``String solitons,''
Phys. Rept. \textbf{259} (1995), 213-326, arXiv:hep-th/9412184 [hep-th].

\bibitem{Reall:2019sah}
H.S.~Reall and J.E.~Santos,
``Higher derivative corrections to Kerr black hole thermodynamics,''
JHEP \textbf{04}, 021 (2019), arXiv:1901.11535 [hep-th].

\bibitem{Gross:1986mw}
D.J.~Gross and J.H.~Sloan,
``The quartic effective action for the heterotic string,''
Nucl. Phys. B \textbf{291} (1987), 41-89.

\bibitem{Bergshoeff:1989de}
E.A.~Bergshoeff and M.~de Roo,
``The quartic effective action of the heterotic string and supersymmetry,''
Nucl. Phys. B \textbf{328} (1989), 439-468.

\bibitem{Bobev:2021oku}
N.~Bobev, A.M.~Charles, K.~Hristov and V.~Reys,
``Higher-derivative supergravity, AdS$_{4}$ holography, and black holes,''
JHEP \textbf{08} (2021), 173.

\bibitem{Ozkan:2013nwa}
M.~Ozkan and Y.~Pang,
``All off-shell $R^{2}$ invariants in five dimensional $\mathcal{N} = 2$ supergravity,''
JHEP \textbf{08} (2013), 042, arXiv:1306.1540 [hep-th].

\bibitem{Ozkan:2013uk}
M.~Ozkan and Y.~Pang,
JHEP \textbf{03} (2013), 158
[erratum: JHEP \textbf{07} (2013), 152, arXiv:1301.6622 [hep-th].

\end{thebibliography}
\end{document}